\documentstyle[preprint,aps,epsfig]{revtex}
\tightenlines
\newcommand{\be}{\begin{equation}}
\newcommand{\ee}{\end{equation}}
\newcommand{\ba}{\begin{eqnarray}}
\newcommand{\ea}{\end{eqnarray}}
\newcommand{\bsigma}{\mbox{\boldmath $\sigma$}}

\newcommand{\bxi}{\mbox{\boldmath $\xi$}}
\newcommand{\bnabla}{\mbox{\boldmath $\nabla$}}

\newcommand{\bPi}{\mbox{\boldmath $\Pi$}}

\newcommand{\br}{\bbox{r}}
\newcommand{\bk}{\bbox{k}}

\newcommand{\bv}{\bbox{v}}
\newcommand{\bw}{\bbox{w}}
\newcommand{\eps}{\epsilon}
\newcommand{\al}{\alpha}
\begin{document}
\title{Randomly Driven Granular Fluids: large scale structure}
\author{
T.P.C. van Noije and M.H. Ernst\\
{\it Instituut voor Theoretische Fysica, Universiteit Utrecht,
Postbus 80006,
3508 TA Utrecht, The Netherlands}\\
E. Trizac\thanks{Present address:
Laboratoire de Physique Th\'eorique et Hautes Energies,
B\^atiment 211, Universit\'e  Paris-Sud,
91405 Orsay Cedex, France. }
and I. Pagonabarraga\\
{\it FOM Institute for Atomic and Molecular Physics, Kruislaan 407, 1098
SJ Amsterdam, The Netherlands}}
\maketitle
\begin{abstract}
The nonequilibrium steady state of a granular fluid, driven by a
random external force, is demonstrated to exhibit long range correlations,
which behave as
$\sim 1/r$ in three and $\sim \ln(L/r)$ in two dimensions.
We calculate the corresponding structure factors
over the whole range of wave numbers, and find
good agreement with two-dimensional molecular dynamics simulations.
It is also shown by means of a mode coupling calculation, how the mean
field values for the steady state temperature and collision frequency, as
obtained from the Enskog-Boltzmann equation, are renormalized by long
wavelength hydrodynamic fluctuations.
\end{abstract}
\narrowtext
\section{Introduction}
Systems of granular particles, like grains of sand or more ideally
glass, plastic
or metal beads, exhibit
different flow regimes \cite{jaeger}, depending on the external forcing.
A systematic experimental study of
the rapid or collisional flow regime as compared to
the quasi-static, slow or frictional regime was first performed by
Bagnold \cite{bagnold} using an annular shear cell.
Later a similar but more refined characterization was made in
Ref.\ \cite{savage+sayed}.
The possibility of coexistence of different flow regimes was observed in
an experimental study of flows down an inclined chute \cite{jackson}.

In several more recent experimental studies of the {\em rapid granular flow}
regime, more microscopic properties have been measured.
In Ref.\ \cite{warr} the
fluidization behavior of a vertically vibrated
two-dimensional model granular material
has been investigated using high speed photography.
Patterns at the surface of a vertically vibrated
granular layer, analogous to Faraday waves in molecular fluids,
have been observed in Ref.\ \cite{swinney} and
stimulated the interest of many theorists \cite{patterns}.
An understanding of these patterns through a derivation of e.g.\
an amplitude equation \cite{hohenberg} 
from the hydrodynamic description of the system,
is still lacking, however.
In Ref.\ \cite{gollub} the effect of inelastic collisions on the
formation of clusters is investigated
in a system of particles rolling on a smooth
surface and driven by a moving wall.
Finally, Ref.\ \cite{olafsen} studies the steady state of a
vertically shaken granular monolayer, and discusses clustering,
inelastic collapse and long range order.

Even rapid flows of {\em model} granular materials are poorly understood in
general, since complicating effects, such as
gravity and interactions with boundaries, have to be taken into
account.
If the model granular material consists of spherical grains with a
smooth surface, collisions between particles can be characterized only
by their coefficient of restitution $\alpha$ or their inelasticity
$\eps=1-\alpha^2$.
We assume this coefficient of restitution to be a constant, independent
of the relative velocity between the colliding particles, and refer to
the model as the {\em inelastic hard sphere} (IHS) model.
Dissipative collisions complicate the dynamics in a nontrivial way:
they may cause the system to
become unstable, and give rise, for instance, to clustering;
they create several new intrinsic length scales that
might interfere for small inelasticity with the system size $L$, and for large
inelasticity with the mean free path $l_0$.

By driving an IHS fluid by boundaries or external fields 
it can reach a steady or
oscillatory state.
Due to the existence of these new ``cooling'' lengths, this
state is frequently inhomogeneous, where the spatial gradients become
larger at higher inelasticity.
Only for small inelasticity the mean free path is well separated from
the scale on which the macroscopic fields vary, and a hydrodynamic
description \cite{jbs} through Navier-Stokes 
or Burnett equations is expected to hold.

In the present paper we investigate the properties of an IHS
fluid
that is heated uniformly so that it reaches a spatially homogeneous
steady state.
This way of forcing, where a random external force accelerates a particle,
was proposed by Williams and MacKintosh
\cite{williams} for inelastic particles moving on a line.
Peng and Ohta \cite{peng} performed simulations on a 2D
version of this model.
In two dimensions the model may be considered to describe
the dynamics of light
disks moving on an air table, a system which has been
investigated experimentally in Ref.\ \cite{oger}.
In three dimensions it can be extended to include gravitational and drag
forces, making it to some extent
relevant for gas-fluidized beds \cite{jackson1}
when hydrodynamic interactions
are unimportant.
A similar IHS model with random external accelerations has been used by
Bizon and Swinney \cite{bizon} in their computer simulations to 
test continuum theories 
for vertically vibrated layers of granular material.

In the present paper we will describe the randomly driven IHS fluid in
two and three dimensions, and characterize its nonequilibrium steady state
(NESS).
The single particle velocity distribution
function in the NESS has been calculated in Ref.\
\cite{granmat} from the Enskog-Boltzmann equation
and was shown to be well approximated by a Maxwellian,
except for an overpopulated tail $\sim\exp(-A c^{3/2})$, where $c$ is
the velocity scaled by the thermal velocity and
$A\sim 1/\sqrt{\eps}$.
Computer simulations of the one-dimensional system
of Ref.\ \cite{williams} showed the existence of long range spatial
correlations in the steady state, which were addressed theoretically in
in Ref.\ \cite{swift}.
Here we will give quantitative predictions for long
range correlations \cite{dorfman} in the two- and three-dimensional NESS. 
Moreover, we extend the mode coupling theory of Brito and Ernst
\cite{brito+ernst} to analyze how long wavelength fluctuations in the NESS
renormalize the
mean field predictions of kinetic theory, and use this theory to
calculate the renormalized temperature and collision frequency in the
NESS.

To obtain an adequate description of the structures
in steady granular flows
one does not only need the equations of fluid dynamics for the average
macroscopic behavior, but also the spatial
correlation functions $G_{ab}(\br)$, and their Fourier transforms, the
structure factors $S_{ab}(\bk)$.
Let $\delta a(\br,t)=a(\br,t)-\langle a(\br,t)\rangle$ with
($a=n,T,u_\alpha$) be the fluctuations of the slowly varying
fields $a(\br,t)$, i.e.\
the local density, $n(\br,t)$, local temperature, $T(\br,t)$, and
local flow velocity $u_\alpha(\br,t)$
($\alpha=x,y,\dots$),
around their average values $\langle
a(\br,t)\rangle$.
Then the objects of interest are the correlation functions in the NESS,
which are given by the limit,
\ba
G_{ab}(\br)=\lim_{t\rightarrow \infty}
\frac{1}{V}\int {\rm d}\br^\prime\langle \delta
a(\br+\br^\prime,t)
 \delta b
(\br^\prime,t)\rangle\nonumber\\
S_{ab}(\bk)=\lim_{t\rightarrow \infty}
V^{-1} \langle \delta a(\bk,t) \delta b(-\bk,t)\rangle.
\label{eq:strfa}
\ea
Here $\langle \dots \rangle$ is an average over some initial distribution,
$\delta a(\bk,t)$ is the spatial Fourier transform of
$\delta a(\br,t)$, and
$S_{ab}(\bk)$ that of $G_{ab}(\br)$.
Moreover we consider the unequal-time correlation functions in the NESS,
defined as
\be
F_{ab}(\bk,t)=\lim_{t^\prime\rightarrow \infty}
V^{-1}
\langle \delta a(\bk,t^\prime+t) \delta
b(-\bk,t^\prime)\rangle,
\label{eq:timec}
\ee
where $F_{nn}(\bk,t)$ is the {\em intermediate scattering function}
\cite{hansen}.
The {\em dynamic structure factor} is then
\be
S_{nn}(\bk,\Omega)={\rm Re} \tilde{F}_{nn}(\bk,z=i \Omega+0),
\label{eq:dynsf}
\ee
where $\tilde{F}_{ab}(\bk,z)$ is the
Laplace transform of $F_{ab}(\bk,t)$.

The paper is organized as follows.
In section \ref{sec:macro} we show how the macroscopic equations for
granular flow are modified to account for the external driving/heating by
the random accelerations.
Section \ref{sec:noise} characterizes the noise of
external and internal fluctuations, and
the structure factors and spatial correlation
functions are calculated in sections \ref{sec:strf} and \ref{sec:analytical}.
The latter section also presents the mode coupling calculations for the
temperature and the collision frequency in the NESS.
Computational details of our molecular dynamics (MD) simulations
are described in section \ref{sec:details},
and section \ref{sec:simul} compares our predictions with simulations.
Some general comments and
conclusions are presented in section \ref{sec:conc}.

\section{Macroscopic Equations}
\label{sec:macro}
Consider a system of inelastic disks or spheres (IHS) ($d=2,3$),
driven by a heat source, which is described as a random acceleration,
$\hat{\bxi}_i$,
\be
\frac{{\rm d} {\bv}_i}{{\rm d}t}=\frac{{\bf F}_i}{m}+\hat{\bxi}_i(t).
\label{eq:ranfo}
\ee
Here ${\bf F}_i$ is the systematic force on particle $i=(1,2,\dots,N)$
due to inelastic collisions.
If the time constant of the heat source is much smaller than the mean
free time $t_0$ between collisions, then $\hat{\bxi}_i(t)$ can be
considered as Gaussian white noise with zero mean and correlation,
\be
\overline{\hat{\xi}_{i\alpha}(t) \hat{\xi}_{j\beta}(t^\prime)}=\xi_0^2
\delta_{ij}\delta_{\alpha\beta} \delta(t-t^\prime),
\label{eq:ranfo2}
\ee
where $\alpha,\beta=\{x,y,\dots\}$ denote Cartesian components of
vectors or tensors.
The over-line indicates an average over the noise source.
It is understood that the ensemble average in Eqs.\ (\ref{eq:strfa}) and
(\ref{eq:timec}),
denoted by the angular brackets,
also includes this noise average.
To guarantee conservation of total momentum the random force has to
obey the constraint $\sum_i \hat{\bxi}_i(t)= \bbox{0}$.
In thermodynamically large systems this constraint gives a correction
to (\ref{eq:ranfo2}) of ${\cal O}(1/N)$, which can be neglected.

The uniformly heated fluid is described by the standard macroscopic
equations of fluid dynamics, where the temperature equation is
supplemented with an additional source term
$m\xi_0^2$, and a sink term $\Gamma$ to account
respectively for the heating and the energy loss
through inelastic collisions:
\ba
&&\partial_t n +  \bnabla\cdot (n \bbox{u})=0\nonumber\\
&&\partial_t \bbox{u}+ \bbox{u}
\cdot
\bnabla \bbox{u}= -\frac{1}{\rho}  \bnabla\cdot  \bPi\nonumber\\
&&\partial_t T + \bbox{u}\cdot  \bnabla T
= -\frac{2}{d n}  (\bnabla\cdot \bbox{J}
+\bPi: \bnabla \bbox{u})
-\Gamma
+ m \xi_0^2,
\label{eq:hydro}
\ea
where $\rho=m n$, ${\bf u}$ the flow velocity, and
$\textstyle{\frac{1}{2}} d nT$ the kinetic energy density in the local
rest frame of the IHS fluid.
The pressure tensor $\Pi_{\alpha\beta}=p \delta_{\alpha\beta}
+\delta\Pi_{\alpha\beta}$ contains the local pressure $p$ and the
dissipative momentum flux
$\delta\Pi_{\alpha\beta}$, which is proportional to $\nabla_\alpha
u_\beta$ and contains the kinematic and longitudinal viscosities $\nu$
and $\nu_l$, defined below Eq.\ (\ref{eq:app1}) of appendix \ref{sec:appA}.
The constitutive relation for the heat flux, $\bbox{J}=-\kappa \bnabla
T$, defines the heat conductivity $\kappa$.
For {\em small} inelasticity the transport coefficients $\nu$, $\nu_l$,
and $\kappa$ are assumed to be given by the Enskog theory for a dense
gas of elastic hard spheres (EHS) \cite{chapman}.

To lowest order in the spatial inhomogeneities the sink term, representing
the energy loss through inelastic collisions, is given by \cite{noije}
\be
\Gamma=2\gamma_0 \omega T.
\label{eq:Gloss}
\ee
It is proportional to the granular
temperature, to the average Enskog collision frequency ($\omega\sim
\sqrt{T}$, explicitly given in Eq.\ (\ref{eq:app2}) of the appendix), and to
the coefficient of inelasticity $\eps=1-\alpha^2\equiv 2 d \gamma_0$,
where $\alpha$ is the coefficient of normal restitution.
To explain the term $m\xi_0^2$ in (\ref{eq:hydro}) we calculate the
energy gain of a single particle due to the random force in a small
time $\delta t$.
This is done by formally integrating (\ref{eq:ranfo}) and averaging
over the noise source, i.e.,
\be
\frac{1}{2}m \left[\overline{v_i^2 (t+\delta
t)}-\overline{v_i^2 (t)}\right]=\frac{d}{2}m \xi_0^2
\delta t,
\ee
where (\ref{eq:ranfo2}) has been used.
Note that we have defined the granular temperature as twice the
average random kinetic energy per
translational degree of freedom, so that the Boltzmann
constant $k_B$ does not appear in its definition.

The above equations provide a consistent description for the heated IHS
fluid at small inelasticity.
The energy is not conserved in inelastic collisions, and consequently
the temperature is {\em not} a {\em hydrodynamic} mode, but a {\em
kinetic} mode with a relaxation rate $\propto \gamma_0 \omega$.
Nevertheless, at small inelasticities ($\gamma_0\ll 1$), it is
consistent to include temperature among the slowly changing
macroscopic variables, which describe the dynamics of the system on
time scales $t$ large compared to the mean free time $t_0=1/\omega$,
and on spatial scales $\lambda$ large compared to the mean free path,
$l_0=v_0 t_0$, where $v_0=\sqrt{2 T/m}$ is the thermal velocity.

At large inelasticities, where $\eps\sim {\cal O}(1)$, the
temperature is a {\em fast} kinetic mode, that decays on the
time scale $t_0$, and cannot be included among the slow macroscopic
variables.
In that case the IHS fluid becomes {\em athermal}, and the slow macroscopic
fields
only involve the density and flow field, as is the case in lattice gas
cellular automata without energy conservation
\cite{fhp,rothman+zaleski}.

Let us consider the decay of temperature in more detail.
For a homogeneous state, the fluid dynamic equations (\ref{eq:hydro})
will have as a  solution  $n(\br,t)=n$, $\bbox{u}(\br,t)=\bbox{0}$, and
$T(\br,t)=T(t)$, the latter satisfying
\be
\partial_t T(t)=-\Gamma+ m\xi_0^2.
\label{eq:Tdec}
\ee
For long times the system approaches a steady state
with a constant temperature, determined by $m \xi_0^2=2
\gamma_0\omega T$.
As $\omega\sim \sqrt{T}$, we obtain
the mean field prediction (\ref{eq:app2}), as deduced from the Enskog
theory,
\be
T_{\rm E} \equiv m \left(\frac{\xi_0^2 \sqrt{\pi}}{2 \gamma_0 \Omega_d \chi n
\sigma^{d-1}}\right)^{2/3}.
\label{eq:Tss}
\ee
Further symbols are defined below (\ref{eq:app1}) in the appendix.
To obtain the final approach to the NESS, we linearize
Eq.\ (\ref{eq:Tdec}) around $T_{\rm E}$ in Eq.\ (\ref{eq:Tss}).
This yields an exponential approach, i.e.\
\be
\delta T(t)\equiv T(t)-T_{\rm E}= \delta T(0) \exp[-3\gamma_0\omega t].
\label{eq:Texp}
\ee
In fact $3\gamma_0 \omega$ can be identified as the
decay rate $z_H(0)$ of the long wavelength components of the
temperature fluctuations, as derived in section \ref{sec:strf} below
(\ref{eq:eigenva}).
The exact time dependent solution of Eq.\ (\ref{eq:Tdec}) can be obtained
implicitly ($t$ as a function of temperature), and reads
\be
f\left(\sqrt{\frac{T(t)}{T_{\rm E}}}\right) - f\left(\sqrt{\frac{T_0}{T_{\rm
E}}}\right) =
-\frac{3}{2} \frac{m \xi_0^2}{T_{\rm E}}\, t,
\label{eq:Tdet}
\ee
where
\be
f(x) = \ln|x-1| -\frac{1}{2}\,\ln(x^2+x+1) +\sqrt{3} \, \hbox{arctan}
\left(\frac{2x+1}{\sqrt{3}}\right).
\label{eq:deff}
\ee
and $T_0$ is the temperature at $t=0$.

\section{Noise characteristics in the NESS}
\label{sec:noise}
The goal of this paper is to analyze the effects of spatial
{\em fluctuations}
$\delta a(\br,t)$ with ($a=n,T,u_\alpha$) around the NESS on
hydrodynamic space and time scales.
As we are dealing with fluctuations, we linearize the nonlinear
equations (\ref{eq:hydro}) around the NESS, with the result
(\ref{eq:app1}) of appendix \ref{sec:appA}.
Moreover, to extend the average equations to fluctuating equations,
valid on mesoscopic spatial and temporal scales, we need to calculate the {\em
external noise} terms $\hat{\bxi}^{\rm ex}(\br,t)$ and $\hat{\theta}^{\rm
ex}(\br,t)$ that contribute to $\partial_t \bbox{u}$ and $\partial_t
T$ in (\ref{eq:app1}).
These terms originate from the random acceleration $\hat{\bxi}_i(t)$,
which enters in the microscopic equations of motion (\ref{eq:ranfo}).
By starting from the microscopic expressions for the momentum and
energy density one finds that the noise sources are given by the {\em long}
wavelength components of
\ba
\hat{\bxi}^{\rm ex}(\br,t)&=&\frac{1}{n}\sum_i \hat{\bxi}_i(t)\delta
(\br-\br_i(t))\nonumber\\
\hat{\theta}^{\rm
ex}(\br,t)&=&\frac{2 m}{d n}\sum_i \bbox{v}_i(t)\cdot \hat{\bxi}_i(t)\delta
(\br-\br_i(t)).
\ea
These fields are again Gaussian white noise with zero mean and
correlations
\ba
\overline{\hat{\xi}_\alpha^{\rm ex}(\br,t)
\hat{\xi}_\beta^{\rm ex}(\br^\prime,t^\prime)}
&=&\frac{1}{n}\xi_0^2 \delta_{\alpha\beta}
\delta(\br-\br^\prime) \delta(t-t^\prime)\nonumber\\
\overline{\hat{\theta}^{\rm ex}(\br,t)
\hat{\theta}^{\rm ex}(\br^\prime,t^\prime)}
&=& \frac{4 m T}{d n}\xi_0^2
\delta(\br-\br^\prime) \delta(t-t^\prime),
\label{eq:extno}
\ea
as follows from (\ref{eq:ranfo2}).

Next, we argue on the basis of the hydrodynamic equations
(\ref{eq:hydro}) that there exists, close to the NESS, a range of
hydrodynamic wave numbers $k\gg k^\ast$, the so called {\em elastic regime},
where the dynamics
of the fluctuations is the same as in a fluid of elastic hard spheres
or disks, and is
driven by {\em internal noise} that will be studied next.
The validity of the hydrodynamic equations (\ref{eq:hydro}) and
(\ref{eq:app2}) is
restricted to wave
numbers $k\ll 2\pi/l_0$ (to guarantee {\em separation} of
kinetic and hydrodynamic scales),
and to $k\ll 2\pi/\sigma$, where $\sigma$ is the disk or sphere diameter
(to guarantee that the Euler equations involve strictly {\em local}
hydrodynamics).
So, for the existence of an elastic regime in the IHS hydrodynamics,
the following constraints must be satisfied
\be
k^\ast \ll k \ll {\rm min}\left\{\frac{2\pi}{l_0},\frac{2\pi}{\sigma}
\right\}.
\label{eq:ineq}
\ee
Moreover, following McNamara \cite{mcnamara} we can distinguish a {\em
dissipative regime}, $k l_0 \ll \eps$ [typically $k l_0\lesssim
{\cal O}(\eps^2)$], and a {\em standard regime}, $k l_0\gg\eps$
[typically $k l_0\gtrsim {\cal O}(\sqrt{\eps})$], separated by a {\em
crossover regime}, around $k l_0 \sim{\cal O}(\eps)$.
In the dissipative regime, dissipation dominates
compression effects and sound propagation,
which are ${\cal O}(k l_0)$, as well as heat conduction,
which is ${\cal O}(k^2 l_0^2)$.
In the standard regime, dissipation effects are of the same order as
heat conduction.
As a consequence, the hydrodynamic modes and
their propagation velocities are those of a fluid of {\em elastic}
particles, while the corresponding damping rates
of heat and sound modes still depend on the
inelasticity.
Only in the {\em elastic regime}, $k l_0\gg k^\ast l_0
\sim{\cal O}(\sqrt{\eps})$,
also these damping coefficients attain their elastic values.
The above argument applies for small enough $\eps=1-\alpha^2=2
d\gamma_0$, where the inequalities (\ref{eq:ineq}) are obeyed,
and an elastic regime exists and is well separated from the dissipative
regime.

In the {\em elastic regime} the equations for the macroscopic deviations from
the NESS are the same as those for a fluid of {\em elastic} hard spheres,
deviating from {\em thermal} equilibrium.
To describe fluctuating mesoscopic hydrodynamics on these length
scales, one can add {\em internal} noise $\hat{\bxi}^{\rm in}(\br,t)$
and $\hat{\theta}^{\rm in}(\br,t)$,
describing the rapid microscopic degrees of freedom.
The noise strength of the internal fluctuations can be obtained from
the fluctuation-dissipation theorem \cite{landau,noije} for the EHS
fluid,
and is given in Fourier representation by
\ba
V^{-1}\overline{\hat{\xi}_\alpha^{\rm in}(\bk,t)\hat{\xi}_\beta^{\rm
in}(-\bk,t^\prime)}&=&\frac{2 T }{\rho}k^2 \left[\nu (\delta_{\alpha\beta}
-\hat{k}_\alpha\hat{k}_\beta)+ \nu_l
\hat{k}_\alpha\hat{k}_\beta\right]\delta(t-t^\prime)\nonumber\\
V^{-1}\overline{\hat{\theta}^{\rm in}(\bk,t)\hat{\theta}^{\rm
in}(-\bk,t^\prime)}&=&\frac{8\kappa T^2}{d^2 n^2} k^2
\delta(t-t^\prime),
\label{eq:intno}
\ea
where $\nu$, $\nu_l$ and $\kappa$ are the transport coefficients for
the EHS fluid, and $\hat{k}_\alpha$ is a component of the unit vector
$\hat{\bk}=\bk/k$.
The {\em effective} noise in the heated IHS fluid may therefore be
described by the sum of external and internal noise,
$\hat{\bxi}(\bk,t)=\hat{\bxi}^{\rm ex}(\bk,t)+\hat{\bxi}^{\rm
in}(\bk,t)$, with a similar expression for $\hat{\theta}(\bk,t)$.
The noise characteristics of $\hat{\bxi}(\bk,t)$ and
$\hat{\theta}(\bk,t)$ {\em interpolate} between two limiting behaviors and
the corresponding noise strengths are given by the sum of
(\ref{eq:extno}) and (\ref{eq:intno}).

Having specified the characteristics of the noise sources in the
macroscopic equations, we conclude this section by summarizing the
Langevin-type equations that describe the dynamics of the slow
fluctuations.
To do so, it is convenient to introduce the Fourier modes $\delta
a(\bk,t)\exp[\imath\bk\cdot\br]$ of the linearized hydrodynamic equations
(\ref{eq:app1}).
The mesoscopic equations, valid on hydrodynamic space and
time scales, i.e.\ $t\gg t_0$ and $k l_0 \ll 1$, then take
the form
\be
\partial_t \delta{\bf a}(\bk,t)= {\bf M}(\bk)
\delta{\bf a}(\bk,t)+\hat{\bf f}(\bk,t),
\label{eq:meso}
\ee
where the components of the vector ${\bf a}$ are labeled
with $a=\{n,T,l,\perp\}$.
Here $a=l$ refers to the longitudinal velocity component
$u_l(\bk,t)=\hat{\bk}\cdot \bbox{u}(\bk,t)$,
and $a=\perp$ refers to $(d-1)$
transverse components of $\bbox{u}(\bk,t)$.
The matrix $M_{ab}$ with $a,b=\{n,T,l,\perp\}$ is given explicitly
in (\ref{eq:app3}), and
$\hat{f}_a(\bk,t)$ is Gaussian white
noise with {\em nonvanishing} components for $a=T,l,\perp$ and
correlation function
\be
V^{-1}\overline{\hat{f}_a(\bk,t)\hat{f}_b(-\bk,t^\prime)}=
C_{ab}(k)\delta(t-t^\prime).
\label{eq:corsca}
\ee
The noise strength $C_{ab}(k)=\delta_{ab}C_{ab}(k)$ is obtained by
taking the Fourier transform of (\ref{eq:extno}) together with
(\ref{eq:intno}), and
is only nonvanishing for the following diagonal elements
\ba
C_{TT}(k)&=&\frac{4 m T\xi_0^2}{d n}+\frac{8
\kappa T^2 k^2}{d^2 n^2}=\frac{4 T \Gamma}{d n}+\frac{8
\kappa T^2 k^2}{d^2 n^2}\nonumber\\
C_{ll}(k)&=&\frac{\xi_0^2}{n}
+\frac{2 \nu_l T k^2}{\rho}=\frac{\Gamma}{\rho}
+\frac{2 \nu_l T k^2}{\rho}\nonumber\\
C_{\perp\perp}(k)&=&\frac{\xi_0^2}{n}
+\frac{2 \nu T k^2}{\rho}=\frac{\Gamma}{\rho}
+\frac{2 \nu T k^2}{\rho},
\label{eq:cnoise}
\ea
where the NESS condition of Eq.\ (\ref{eq:Tdec}), i.e.\ $\Gamma=2\gamma_0\omega
T=m\xi_0^2$, has been used.

\section{Structure factors}
\label{sec:strf}
The equal-time structure factors, introduced in
(\ref{eq:strfa}),
obey the equations of motion
\be
\partial_t S_{ab}(\bk)= \sum_c\left\{M_{ac}(\bk) S_{cb}(\bk)
+M_{bc}(-\bk) S_{ac}(\bk)\right\}+C_{ab}(k),
\label{eq:eqmosf}
\ee
which follows by formally integrating (\ref{eq:meso}) and using
(\ref{eq:corsca}).
The left hand side of (\ref{eq:eqmosf}) vanishes since
the structure factors do not depend on time in the NESS.
The resulting equation can be solved by spectral
analysis, or numerically.
The spectral analysis is summarized in appendix \ref{sec:appA},
where $w_{\lambda a}$
and $v_{\lambda a}$ are respectively the $a$-th component of the right
and left eigenvectors of the hydrodynamic matrix ${\bf M}$, and
$z_\lambda(k)$ is the corresponding eigenvalue.

Taking then the scalar product of (\ref{eq:eqmosf}) on both sides with left
eigenvectors (\ref{eq:ev}) of the appendix yields
\be
\sum_{a b}\langle v_{\lambda a}(\bk)|S_{a b}(\bk)|v_{\mu b}
(-\bk)\rangle=-\sum_{a b}\frac{\langle
v_{\lambda a}(\bk)|C_{a b}(k)|v_{\mu b}(-\bk)\rangle}{z_\lambda(k)+
z_\mu(k)}.
\ee
Using the completeness relation (\ref{eq:compl}) and the fact that off-diagonal
elements
of $C_{ab}$ in (\ref{eq:cnoise}) vanish, we obtain
\be
S_{ab}(\bk)=-\sum_{\lambda\mu c}\frac{w_{\lambda a}(\bk) v_{\lambda
c}(\bk) C_{cc}(k) v_{\mu c}(-\bk) w_{\mu b}(-\bk)}{z_\lambda(k)+
z_\mu(k)},
\label{eq:staticsf}
\ee
which is the final result for the static structure factors.
The time correlation function (\ref{eq:timec}) in the NESS reduces in a
similar manner to
\be
F_{ab}(\bk,t)
=\sum_{\lambda c} \exp[z_\lambda(k)t] w_{\lambda a}(\bk) v_{\lambda
c}(\bk) S_{c b}(\bk).
\label{eq:timesf}
\ee
The above result corresponds to the Landau-Placzek theory \cite{hansen}
for hydrodynamic correlations in the NESS.

To obtain more explicit results we need the explicit forms of eigenvalues and
eigenvectors,
which have only been calculated for small $k$.
The eigenvalue equation can be solved numerically for any given wave
number, and the results are illustrated in Fig.\ \ref{fig:Stheo} for the
two-dimensional case.
Transport coefficients, equation of state $p(n,T)$, 
and pair correlation function
at contact are obtained from
Enskog's theory for {\em elastic} hard spheres or disks.
The generic features of the spectrum in Fig.\ \ref{fig:Stheo} are the same as
McNamara's case ``$\alpha=\beta=0$'', illustrated in Fig.\ 6(a) of his
study \cite{mcnamara} on hydrodynamics of granular materials, which
corresponds to a temperature and density independent heat source.
However, neither the equation of state, nor the transport coefficients,
used in Ref.\ \cite{mcnamara}, correspond to the heated fluid of
inelastic hard spheres, used in the present simulations.
In the hydrodynamic regime ($k l_0\lesssim 1$) all eigenvalues are found to
be {\em negative} for nonvanishing wave numbers (see appendix
\ref{sec:appA}).
So, all modes are linearly {\em stable}.

With the help of Mathematica the structure factors in the steady state
have been calculated numerically from Eq.\ (\ref{eq:eqmosf}) with 
$\partial_t S_{ab}(\bk)=0$ for a given wave number.
The resulting structure factors are shown by solid lines in Fig.\
\ref{fig:int-ext}, and will be tested against MD simulations in section
\ref{sec:simul}.

Next, we present analytic results for the dissipative regime ($k l_0\ll
\gamma_0$).
The eigenvalues on the largest spatial scales can be determined as an
expansion in powers of $k$ at a fixed value of $\gamma_0$, with the
results
\ba
z_\perp(k)&=&-\nu k^2\nonumber\\
z_\pm(k)&=& \mp i k v_D - D_S k^2\nonumber\\
z_H(k)&=&-3 \gamma_0 \omega+ D_H k^2.
\label{eq:eigenva}
\ea
In later applications the explicit form of the eigenvectors of ${\bf
M}(\bk)$ is needed to lowest order in $\bk$.
They read
\ba
\bw_\perp(\bk)=(0,0,0,1)\,\,\,\,\,\,\,\,\,\,\,\,\,\,\,\,\,\,\,\,\,\,\,\,\,\,\,
\,\,\,\,\,\,\,\,\,\,\,\,\,\,\,\,\,\,\,\,\,\,\,\,\,\,\,\,\,
&&\bv_\perp(\bk)=(0,0,0,1)\nonumber\\
\bw_\pm(\bk)=\frac{1}{\sqrt{2}}(1,- g(n)T /3 n,\pm v_D/n,0)
\,\,\,\,\,\,\,\,\,&&\bv_\pm(\bk)=
\frac{1}{\sqrt{2}}(1,0,\pm n/v_D,0)\nonumber\\
\bw_H(\bk)=(0,1,0,0)\,\,\,\,\,\,\,\,\,\,\,\,\,\,\,\,\,\,\,\,\,\,\,\,\,\,
\,\,\,\,\,\,\,\,\,\,\,\,\,\,\,\,\,\,\,\,\,\,\,\,\,\,\,\,\,
&&\bv_H(\bk)=( g(n)T/3 n,1,0,0).
\label{eq:eigenve}
\ea
The coefficients $g(n)$, $v_D$, and $D_\lambda$ with
$\lambda=\{\perp,H,\pm\}$ are calculated in the appendix.
In the dissipative regime there are two propagating sound modes
($\lambda=\pm$) with a propagation speed $v_D$ and a
damping constant $D_S$.
There is a {\em kinetic} heat mode ($\lambda=H$), with a long wavelength
relaxation rate $z_H(0)=-3 \gamma_0 \omega$, in agreement
with Eq.\ (\ref{eq:Texp}).
Therefore, on the largest spatial scales the temperature deviations have
decayed to
zero, and temperature gradients do not exist; there is no heat
conduction.
In addition, there are $(d-1)$ transverse velocity or shear modes
($\lambda=\perp$), which are purely diffusive.
The corresponding diffusivity, $D_\perp=\nu$,
has the same form as for EHS.
In (\ref{eq:vD})---(\ref{eq:DH}) the coefficients are expressed
explicitly in terms of thermodynamic quantities and transport
coefficients.

In the standard regime, $k l_0\gg \gamma_0$ (and $\gamma_0$ small)
the eigenvalues for shear and sound modes
are to leading nonvanishing
order the same as for the EHS fluid, where the sound waves propagate with
the {\em adiabatic} sound speed $v_S$ of the elastic fluid, which is
larger than the propagation speed $v_D$ in (\ref{eq:vD}).
The damping of the sound and heat modes, on the other hand,
are larger than in the
elastic fluid due to the inelastic collisions.
In the elastic regime, defined in section \ref{sec:noise},
where $k l_0\gg \sqrt{\gamma_0}$,
all transport coefficients are equal to their EHS values.

In summary, the eigenvalue spectrum $z_\lambda(k)$ for the uniformly
heated IHS fluid is quite different from that of the freely evolving IHS
fluid, linearized around the homogeneous cooling state
\cite{mcnamara}.
In the {\em free} case, all shear modes ($\lambda=\perp$) and the heat
mode ($\lambda=H$) are unstable in the dissipative regime (small $k$),
and propagating modes do not exist for $k l_0\ll \gamma_0$.
Moreover, there exist in this regime a stable diffusive density mode and
a kinetic temperature mode, which combine into two propagating modes for
$k l_0\sim {\cal O}(\gamma_0)$, where crossover occurs from the dissipative to
the standard regime.
In the {\em heated} case, however, all modes are linearly stable, and the
sound modes remain propagating in the dissipative regime down to $k=0$.

\section{Effects of Long Range Correlations}
\label{sec:analytical}
In this section we study static and dynamic structure factors and
corresponding correlation functions at the largest spatial scales.
Moreover, we show by means of a mode coupling calculation, how
average properties, which were calculated in section \ref{sec:macro}
on the basis of a mean field theory (i.e.\ the Enskog-Boltzmann equation)
are renormalized by spatial fluctuations.

The static structure factors have been calculated in
(\ref{eq:staticsf}).
In the {\em dissipative regime} ($k l_0\lesssim \gamma_0$),
the relevant eigenvalues
(\ref{eq:eigenva}) and eigenmodes (\ref{eq:eigenve}) are discussed
below
(\ref{eq:eigenve}).
The dominant singularity at small wave number
of the structure factors
$S_{ab}(k)\sim {\cal O}(1/k^2)$ in (\ref{eq:staticsf}) originates from
pairs of transverse modes, where $2 z_\perp(k)=-2 \nu k^2$, and
from anti-parallel sound modes, where
$z_+(k)+z_{-}(k)=-2 D_S k^2$.
We start with the transverse structure factor, where
only the shear modes in (\ref{eq:eigenve})
contribute, and deduce from the equations above,
\be
S_\perp(k)\equiv S_{\perp\perp}(k)\simeq\frac{\Gamma}{2
\rho\nu k^2},
\label{eq:sperp}
\ee
where the relation (\ref{eq:cnoise}) has been used for $k
l_0\ll\gamma_0$.
The structure factors $S_{ab}(k)$ for
$a,b\neq \perp$ derive their dominant small-$k$ behavior from two
anti-parallel sound modes and we obtain with the help of
(\ref{eq:staticsf}), (\ref{eq:eigenva}), (\ref{eq:eigenve}), and
(\ref{eq:cnoise}) at small $k$,
\be
S_{\parallel}(k)\equiv S_{ll}(k)\simeq \frac{\Gamma}{4\rho D_S k^2},
\label{eq:sparallel}
\ee
where the sound damping constant
in the dissipative regime $D_S$
has been calculated in (\ref{eq:Ds}).
It depends
on the inelasticity.
The contribution of the internal noise is subdominant in this regime.
In a similar manner we find
\be
S_{ab}(k)\simeq B_{ab} S_{ll}(k)
\,\,\,\,\,\,\,\,\,\,\,\,\,\,\,(k\rightarrow
0),
\label{eq:bleue}
\ee
where all nonvanishing coefficients labeled $(ab)=(ll,nn,TT,nT)$ are
listed in Table \ref{bleue}.
The remaining structure factors are of ${\cal O}(1)$ as $k\rightarrow
0$.

Next, we consider the spatial correlation functions $G_{ab}(r)$, which
are the inverse Fourier transforms of $S_{ab}(k)$.
The {\em small}-$k$ behavior of $S_{ab}(k)$, obtained above,
enables us to calculate the {\em
large} $r$ behavior of the spatial correlation functions.
The calculations are given in appendix \ref{sec:appB}.
One finds for the leading large $r$ behavior in {\em three}-dimensional
systems,
\ba
G_\parallel(r) &\simeq& \left(\frac{\Gamma}{8 \pi \rho
\nu}\right)\frac{1}{r}
\nonumber\\
G_\perp(r) &\simeq& \frac{\Gamma}{16 \pi \rho
}\left(\frac{1}{\nu}+\frac{1}
{2 D_S}\right)\frac{1}{r},
\label{eq:large3}
\ea
and in {\em two}-dimensional systems,
\be
G_\parallel(r)\simeq G_\perp(r)\simeq\frac{\Gamma}{8\pi \rho}
\left(\frac{1}{\nu}+\frac{1}{2 D_S}\right)
\ln\left(\frac{L}{r}\right),
\label{eq:large}
\ee
valid for $r\ll L$, where $L$ is the linear dimension of the system.
The subleading large $r$ corrections to (\ref{eq:large}) are constant
terms, independent of $r$.
In the calculations given in appendix \ref{sec:appB},
these constants depend on a
cutoff wave vector $k_{\rm min}=2\pi/L$, used to evaluate the divergent
$\bk$ integrals occurring in the Fourier inversion of $S_{ab}(k)$.
To calculate their precise values the subleading small-$k$ corrections
to (\ref{eq:sperp}) and (\ref{eq:sparallel}) are required.

The long wavelength behavior of the
time dependent correlation function $F_{ab}(k,t)$
in (\ref{eq:timesf}) can be evaluated in a similar manner.
We quote the result in terms of the Laplace transform,
$\tilde{F}_{ab}(k,z)$,
from which the dynamic structure factor (\ref{eq:dynsf}) follows.
In the {\em dissipative regime} ($k l_0\ll\gamma_0$) we find to
leading order for small wave numbers,
\ba
\tilde{F}_{\perp\perp}(k,z)&\simeq&\frac{S_{\perp\perp}(k)}{z+\nu
k^2}\nonumber\\
\tilde{F}_{ll}(k,z)&\simeq&{\textstyle{\frac{1}{2}}}
\sum_{\lambda=\pm}\frac{S_{ll}(k)}{z+ i \lambda k v_D + D_S k^2}.
\ea
In a similar manner we obtain
\be
\tilde{F}_{ab}(k,z)\simeq B_{ab}\tilde{F}_{ll}(k,z),
\ee
where all nonvanishing coefficients $B_{ab}$ are listed in Table
\ref{bleue}.
The dynamic structure factor (\ref{eq:dynsf}) then becomes
\be
S_{nn}(k,\Omega)\simeq  {\textstyle{\frac{1}{2}}}\sum_{\mu=\pm} \frac{ D_S
k^2 S_{nn}(k)}{(\Omega+ \mu k v_D)^2+D_S^2 k^4}.
\ee
It contains only Brillouin peaks, coming from the sound modes.
There is no central Rayleigh peak, because the heat mode is not a slow,
but a fast kinetic mode in this regime.
In the elastic regime $S_{nn}(k,\Omega)$ has the standard Rayleigh and
Brillouin lines of the EHS fluid.

The existence of long range spatial correlations shows that the NESS is
quite different from a thermal equilibrium state \cite{dorfman}.
In fact, the spatial fluctuations also modify (``renormalize'') the mean
field predictions for the averages and the particle distribution
functions.
In appendix \ref{sec:appC} a mode coupling calculation is presented to estimate
the renormalization effects on the average energy per particle
$E=(1/N)\sum_i
\langle \textstyle{\frac{1}{2}} m v_i^2\rangle$ and average collision
frequency $\omega$ in the NESS, and we recall their mean field values, i.e.\
$E_{\rm E}=\textstyle{\frac{1}{2}} d T_{\rm E}$ and $\omega_{\rm E}$
given by (\ref{eq:app2}), i.e.\ $\omega_{\rm E}\propto n
\chi(n)\sqrt{T_{\rm E}}$, where $T_{\rm E}$ is given in
(\ref{eq:Tss}).

As it turns out, the fluctuation contributions, $\delta E$ and $\delta
\omega$, are finite and well-behaved in {\em three} dimensions, but
logarithmically divergent in the system size $L$ in {\em two} dimensions,
so that $d=2$ is the upper critical dimension.
The mode coupling calculations of appendix \ref{sec:appC} yield then in two
dimensions, for large $L$,
\ba
T_{\rm NESS} &\simeq& T_{\rm E} + \frac{{\cal C}_E}{4\pi}
\ln{\left(\frac{\gamma_0 L}{l_0}\right)}\nonumber\\
\omega_{\rm NESS} &\simeq& \omega_{\rm E} + \frac{{\cal C}_\omega}{4\pi}
\ln{\left(\frac{\gamma_0 L}{l
_0}\right)},
\label{eq:eoness}
\ea
where ${\cal C}_E$ and ${\cal C}_\omega$ are calculated in
appendix \ref{sec:appC}.
The argument of the logarithm $\gamma_0 L/l_0$ is an estimate for
the ratio of the values for the right ($k \sim \gamma_0/l_0$) and left
($k_{\rm
min}=2\pi/L$) boundaries of the dissipative $k$ range, where the
small-$k$ behavior in (\ref{eq:sperp}) to (\ref{eq:bleue}) is valid.
The logarithmic correction becomes only appreciable for large systems
with a size $L$, much larger that the so called homogeneous cooling
length $l_T=l_0/\gamma_0$, which diverges in
the elastic limit.
Then the renormalization corrections for small inelasticity vanish as
$\delta T \sim \eps \ln(\eps
L/l_0)$ and $\delta \omega \sim \eps^2 \ln(\eps
L/l_0)$.
Here we have used the relations ${\cal C}_{\rm E}\sim\eps$ and ${\cal
C}_\omega \sim \eps^2$ for $\eps\rightarrow 0$, as can be deduced from the
results in appendix \ref{sec:appC}.
A similar mode coupling theory has been recently used
in Ref.\
\cite{brito+ernst} for freely evolving granular fluids to calculate
the long time decay of the energy, which
deviates from Haff's cooling law due to inhomogeneities
in the hydrodynamic fields, and good agreement between theory and
simulations was found.

Before concluding this section we compare the theoretical predictions
for the structure factors, with and without internal noise in Eqs.\
(\ref{eq:cnoise}) and (\ref{eq:eqmosf}), as shown in Fig.\ \ref{fig:int-ext}.
Inspection of (\ref{eq:sperp}) and (\ref{eq:sparallel}) shows that only
the external noise determines their dominant small-$k$ behavior.
The question then arises what are the effects of internal noise, and is
it meaningful to include it in the theoretical description?
The answer is affirmative, as we will show below.
The steady state solution of (\ref{eq:eqmosf}) {\em without internal
noise} clearly behaves at small wavenumbers as $k^{-2}$,
but the $k$ independent plateau values,
shown by the full theory
(solid lines in Fig.\ \ref{fig:int-ext}), are missing.
These plateau values represent a very short distance correlation $\sim
\delta(\br)$.
Calculation of the plateau value for $S_{\alpha\beta}(\bk)$ yields
$(1/n^2
V)\langle \sum_i v_{i\alpha}v_{i\beta}\rangle=(T/\rho)\delta_{\alpha\beta}$,
i.e.\ the self-correlation term ($i=j$) in the definition
(\ref{eq:strfa}) of $S_{\alpha\beta}(\bk)$, or more explicitly in Eq.\
(\ref{eq:Smomentum}) below.
Then, addition of this plateau value to the numerical solution of
(\ref{eq:eqmosf}) {\em without} internal noise yields the structure
factors $S^{\rm without}_{\alpha\beta}(\bk)$,
shown as dotted lines
in Fig.\ \ref{fig:int-ext}.
For the transverse velocity structure factor the dotted line coincides
with the solid line in Fig.\ \ref{fig:int-ext}, as can be shown
analytically from (\ref{eq:cnoise}) and (\ref{eq:eqmosf}).
Only the structure factor for the longitudinal flow field,
$S_\parallel^{\rm without}(k)$ differs appreciably from
$S_\parallel(k)$ in the relevant intermediate regime,
$0.2\lesssim k\sigma \lesssim 0.5$.

In section \ref{sec:simul} the simulation results will be compared with
the theoretical predictions with and without internal noise.
As it turns out, the comparison shows convincingly that the theory
with/without internal noise agrees/disagrees with the simulations.

\section{Computational details}
\label{sec:details}
In the two subsequent sections
we describe molecular dynamics
simulations performed to verify our theoretical predictions.
In this section we present the computational details,
before testing our theoretical results against simulations in the next
section.

The model studied here has been extensively used in computer
simulations
for the freely evolving case (no forcing) \cite{goldhirsch,McNY,DeB,noije},
as well as for the randomly accelerated case in one \cite{williams} and two
dimensions \cite{peng}.
We consider a system of $N$ inelastic hard disks having diameter
$\sigma$ in a two-dimensional square cell of length $L$,
with periodic boundary conditions. The disks interact via
inelastic collisions with coefficient of normal restitution $\alpha$.
For a colliding pair ($i,j$) of particles having equal masses the
postcollision velocities are:
\begin{eqnarray}
\bv^\ast_i& =& \bv_i - \frac{1}{2}(1+\alpha)
(\bv_{ij}\cdot \hat{\bsigma}) \hat{\bsigma} \nonumber\\
\bv^\ast_j& =& \bv_j + \frac{1}{2}(1+\alpha)
(\bv_{ij} \cdot \hat{\bsigma})\hat{\bsigma} ,
\label{eq:collrule}
\end{eqnarray}
where $\bv_{ij}=\bv_i-\bv_j$, the asterisk
denotes velocities after collision, and
$\hat{\bsigma}$ is a unit vector along the line connecting the
centers of
particle $j$ and particle $i$.

The energy loss in consecutive collisions, which is proportional to
$\eps=1-\al^2$, is compensated by a periodic (in time) and
instantaneous perturbation of all velocities by a
random amount. After every time step $\Delta t$, the velocity of each
particle is modified according to
\be
\bv_i \rightarrow \bv_i +\bbox{\varphi}_i, \qquad 1 \leq i \leq N,
\ee
where the components of the vectors
$\bbox{\varphi}_i$ are taken from a random distribution
of zero mean and variance $\varphi_0$ (in practice
a Gaussian or a flat function of finite support).
The time step $\Delta t$ of this ``heating'' or ``kicking''
is chosen much smaller
than the mean time between successive collisions
of a tagged particle (typically a factor $10^4$ smaller), so
that the system under scrutiny reduces to that described in section
\ref{sec:macro}.
The opposite limit, where $\Delta t$ is much bigger or comparable to
the mean free path, was considered in Ref.\ \cite{puglisi} (in
the presence of an additional external damping or drag force), and in that
limit clustering was observed.
The relation between $\varphi_0$ and $\xi_0$,
the variance of the noise term in Eq.\ (\ref{eq:ranfo}), can be
deduced from the energy fed into the system by the kicks.
This yields straightforwardly,
\be
\xi_0^2 = \frac{\varphi_0^2}{\Delta t}.
\ee
Between the heating events, the motion of the disks is free,
which enables us to implement an event-driven molecular
dynamics scheme with a linked-list method \cite{AT}.
The CPU time however scales like $N^2$ because the lists
in all cells need to be updated after each heating event.

Four parameters determine the state of the
system: the inelasticity $\eps=1-\alpha^2$, the packing fraction
$\phi = \pi N \sigma^2 /(4 L^2)$, the reduced lower wave number cutoff
$k_{\rm min} \sigma = 2\pi \sigma/L$,
and the heating rate $\xi_0^2$.
The values of $N$ investigated in this article vary between
$10^3$ and $10^4$, and we shall restrict our attention to high
packing fractions for which the use of linked lists
implies the most significant reduction of computer time.
We consider the cases of moderate inelasticities ($0.6< \alpha <1$)
and complete inelasticity ($\alpha=0$). For the latter case,
inelastic collapse occurs,
i.e.\ the collision frequency involving only a small
number of correlated particles diverges,
as observed first by McNamara and Young \cite{McNY}
in freely evolving fluids of inelastic hard disks.
Also in our system, for high enough inelasticity
the heating seems never
sufficient to prevent the inelastic collapse.
For $\alpha < 0.5$,
the inelastic collapse has been avoided by introducing
a slight modification of collision rule (\ref{eq:collrule}),
as proposed in \cite{DeB}: in each collision,
the velocities are first computed according
to the standard procedure $(\bv_1,\bv_2) \to (\bv^\ast_1,\bv^\ast_2)$;
the relative velocity $\bv^\ast_{12}$ is then
rotated by a  random angle smaller than a maximum value
$\Theta$
(typically less than a few degrees),
keeping the center of mass velocity fixed. Note that this modified
collision rule does not change the total energy loss
of the colliding pair, and does not introduce any spurious
drag or forcing on the particles.

The structure factors in the NESS have been computed for wave vectors
compatible with the periodic boundary conditions, i.e.\
of the form $(2\pi/L) (n_x,n_y)$. We have obtained the density-density
structure factors
\be
S_{nn}(k) = \frac{1}{V} \biggl\langle \sum_{i,j} \exp(-\imath\, \bk \!\cdot\!
\br_{ij})\biggr\rangle = \frac{1}{V} \biggl\langle \biggl|\sum_i \exp
(-\imath\, \bk\!\cdot\!\br_i)\biggr|^2\biggr\rangle,
\label{eq:Snn}
\ee
and the velocity-velocity structure factor, defined as,
\be
n^2 S_{\alpha\beta}(\bk)=\frac{1}{V}
\biggl\langle \sum_{i,j} v_{i\alpha}v_{j\beta}\, \exp(-\imath\, \bk \!\cdot\!
\br_{ij})\biggr\rangle.
\label{eq:Smomentum}
\ee
Here the averages are taken in the spatially uniform NESS.
The fluctuation $\delta g_\alpha$ in the momentum density,
$\delta g_\alpha(\bk) =\sum_i m v_{i\alpha} \exp(-\imath \bk \cdot \br_i)$,
and those in the flow field are related as $\delta g_\alpha =
\rho \delta u_\alpha$, where $\rho$ is the average mass density in the steady
state.
The second rank tensor $S_{\alpha\beta}(\bk)$ is isotropic and can be
split into a longitudinal and transverse part,
\be
S_{\alpha\beta}(\bk)=\hat{k}_\alpha \hat{k}_\beta S_\parallel(k) +
(\delta_{\alpha\beta}-\hat{k}_\alpha \hat{k}_\beta) S_\perp(k),
\label{eq:decomp}
\ee
where $\hat{k}=\bk/k$.
From Eq.\ (\ref{eq:Snn}), it appears that the
knowledge of $S_{nn}$ requires the computation of an ${\cal O}(N)$
quantity. We can rewrite the velocity-velocity structure factor
so that its
computation also increases linearly with the number of particles. For
example, for the longitudinal part of $S_{\alpha\beta}(\bk)$ in
(\ref{eq:decomp}) we have,
\be
n^2 S_\parallel(k)=\frac{1}{V} \biggl\langle \biggl|\sum_i
\left(\bv_i\cdot\hat{\bk}\right)\,
\exp(-\imath\, \bk\!\cdot\!\br_i)\biggr|^2\biggr\rangle.
\ee
In practice, the different $S_{aa}(k)$ have been computed for every $\bk$
lying in the disk $|\bk|<k_{\hbox{\scriptsize max}} = 6 \pi / \sigma$,
then averaged over shells of thickness $k_{\hbox{\scriptsize min}}=
2\pi/L$, to achieve better accuracy. Moreover, the statistics in the
NESS
has been increased by averaging over time.
Note that the above procedure, which gives insight into the microscopic
to large scale
structure of the system, does not require the knowledge
of the hydrodynamic (coarse-grained) density and velocity fields.

\section{Simulation results}
\label{sec:simul}

\subsection{Approach and characterization of the NESS}

Before addressing the question of the large scale
structure of the inelastic fluid,
we investigate the validity of the macroscopic description given
in section \ref{sec:macro} and \ref{sec:analytical}.
The former section gives the mean field results for the steady state
temperature $T_{\rm E}$ in (\ref{eq:Tss}) and collision frequency
$\omega_{\rm E}$ in (\ref{eq:app2}), based on the Enskog-Boltzmann
equation.
The latter section and appendix C show how the long range spatial
fluctuations renormalize these mean field values and lead to estimates
in (\ref{eq:eoness}) for the corrections $\delta T= T-T_{\rm E}$ and
$\delta \omega=\omega-\omega_{\rm E}$, using a mode coupling
calculation.

When the system is initially prepared
in a configuration having a temperature $T_0$ different from the
steady state temperature $T_{\rm E}$, the time dependence predicted
by the mean field result (\ref{eq:Tdet}) is in good agreement
with the numerical data.
This is shown in Fig.\ \ref{fig:Tdet} for a system with small inelasticity.
When the initial temperature $T_0$ is much larger than $T_{\rm E}$,
the heating at short times is dominated by the inelastic
dissipation, and Eq.\ (\ref{eq:Tdet}) becomes Haff's homogeneous
cooling law for a freely evolving system
\be
\frac{T(t)}{T_0} = \frac{1}{(1+\gamma_0 t /t_0)^2},
\label{eq:homco}
\ee
where $t_0=1/\omega_{\rm E}(T_0)$ is the mean free time in the initial
state.
This can be seen from the asymptotic expansion of the function $f$
defined in (\ref{eq:deff}),
\be
f(x) \simeq \sqrt{3}\, \frac{\pi}{2} - \frac{3}{x} \qquad \hbox{for} \qquad
x\to \infty.
\label{eq:homcool}
\ee
These analytic results for short times are confirmed by MD
simulations, as shown in Fig.\ \ref{fig:Tdetlog}.
Moreover, for initial temperatures $T_0\ll T_{\rm E}$, Eq.\ (\ref{eq:Tdet})
predicts at short times a linear increase of $T$, as in a heated fluid of
elastic hard spheres.
Our simulations confirm this behavior.

Fig.\ \ref{fig:scale} shows that the
measured kinetic energy per particle $T$ is larger than the
temperature
$T_{\rm E}$ predicted on the basis of mean field theory.
This effect is noticeable albeit small in the results
reported in Figs.\ \ref{fig:Tdet} and \ref{fig:Tdetlog},
which correspond to the nearly elastic limit. For the densities studied
here,
we observe (see Fig.\ \ref{fig:scale}) that the
correction $\delta T$ is positive and decreases with decreasing
inelasticity.
The positive excess in temperature
is already present for small inelasticity (see e.g.\ Fig.\
\ref{fig:Tdetlog}), and vanishes as $\eps \rightarrow 0$.
Note that the above results are at variance with those
reported by Peng and Ohta \cite{peng}, who find that
$T/T_{\rm E}$ does not depend on $\alpha$.
The measured collision frequency
$\omega$ is also larger than the Enskog estimate $\omega_{\rm E}$,
as shown
in Fig.\ \ref{fig:scale}.
The excess $\delta \omega$ increases with increasing inelasticity.

We first observe that the simulation data in Fig.\ \ref{fig:scale},
where {\em both} $\omega>\omega_{\rm E}$ and $T> T_{\rm E}$, cannot be
explained consistently as a possible over- or underestimation of the IHS
pair correlation function $\chi_{\rm IHS}$ at contact  
by its value for elastic disks.
On the basis of (\ref{eq:Tss}) and (\ref{eq:app2}) we note that an
overestimation of $\chi_{\rm IHS}$ would increase the collision frequency
(\ref{eq:app2}), and decrease the temperature (\ref{eq:Tss}).
These trends are at variance with the observations.
The discrepancy between the measured and predicted
temperatures and collision frequencies is more likely
due to large scale fluctuations, at least for not too large
inelasticities ($\alpha
\gtrsim 0.6$).
These long range spatial fluctuations renormalize the mean field Enskog
values for the temperature $T_{\rm E}$ and collision frequency
$\omega_{\rm E}$ by amounts $\delta T$ and $\delta \omega$.
The theoretical estimates (\ref{eq:eoness}), based on mode coupling
arguments,
show the correct trends for the dependence of
these corrections on the inelasticity, i.e.\ $\delta T\sim \eps$ and
$\delta \omega \sim \eps^2$ as $\eps\rightarrow 0$, and 
give a rough estimate of the
magnitude of these terms as
illustrated in the inset of Fig.\ \ref{fig:scale}.
We also point out that the system size considered in Fig.\
\ref{fig:scale} is much too small for the asymptotic theory 
(\ref{eq:eoness}) to be
applicable.
For instance at $\alpha=0.8$, we deduce from the data ($l_0 = 1.1 \sigma,
L=60 \sigma$) in the caption of Fig. \ref{fig:scale} that $\gamma_0 L
/l_0\simeq 4.9$.
Consequently, the leading asymptotic term $\ln(\gamma_0 L/l_0)\simeq 1.6$
does not dominate the full mode coupling contribution (\ref{eq:hness}) in
appendix \ref{sec:appC}, where subleading terms of ${\cal O}(1)$ have
been neglected.
The corresponding ratios $\gamma_0 L/l_0$ for Figs.\ \ref{fig:S0.92}
($\alpha=0.92$),
\ref{fig:S0.6hydro} ($\alpha=0.6$) and \ref{fig:S0hydro0.76} ($\alpha=0$)
are respectively 
46, 127 and 297.
This predicts for the systems in Figs.\ \ref{fig:S0.92},
\ref{fig:S0.6hydro} and \ref{fig:S0hydro0.76} 
respectively $T_{\rm NESS}/T_{\rm E} \simeq
1.07$, 1.41 and 1.77, whereas the simulations yield for the observed values
$T/T_{\rm E}\simeq 1.05$, 1.45 and 1.5, which agrees quite well.
The good agreement at $\alpha=0$ is unexpected, as the theory is
constructed under conditions that apply at small inelasticity.
However, the renormalized values for the collision frequency, as
predicted by the mode coupling theory, are much too small.
We find in the above Figs. \ref{fig:S0.92},
\ref{fig:S0.6hydro} and \ref{fig:S0hydro0.76} 
respectively for the theoretical values
$\omega_{\rm NESS}/\omega_{\rm E} \simeq 1.003$, 1.06 and 1.12, whereas the
simulations yield $\omega/\omega_{\rm E}\simeq$ 1.05, 1.36 and 22.9.

For large inelasticity ($\alpha<0.5$), the temperature $T$ and collision 
frequency
$\omega$ depend on the maximum angle $\Theta$
of the random rotations, used to avoid the collapse singularity,
as explained in section \ref{sec:details}.
In fact, $\omega$
diverges at $\Theta=0$ (inelastic collapse), 
in agreement with the
observations of Peng and Ohta \cite{peng}.
Intuitively, one might expect that a {\em larger} randomization of the
postcollision velocities (larger $\Theta$), would more effectively
destroy the correlations leading to inelastic collapse, and
consequently would {\em decrease} the deviations in $T$ and $\omega$
from the mean field and the mode coupling predictions.
This happens indeed for the temperature, but not for the collision
frequency.
At packing fraction $\phi=0.07$, the collision frequency is indeed
monotonically decreasing with increasing $\Theta$ 
(see Table \ref{rouge}).
However, at $\phi=0.2$ it diverges at $\Theta=0$, decreases to a
finite value at $\Theta=5^o$, and increases again to its value at 
$\Theta=10^o$ (see Fig.\ \ref{fig:scale}).
We have no explanation for this behavior.

\subsection{Fluctuations in the NESS}

In this section we analyze the effects of inelasticity on the large
distance behavior of the fluid.
We have computed the structure factors in the spatially homogeneous
NESS of the
inelastic hard disk fluid as
explained in the previous section, and have focused either on values of
$\alpha$ close to the elastic limit, where the theoretical description
is supposed to apply, or on values close to 0, in order to test how large
deviations from the theory might be.
Local mean field values, like $T_{\rm E}$ and $\omega_{\rm E}$, reach
their steady state values rapidly.
However the time scale needed for the structure factors $S_{ab}(\bk)$
and the contributions of spatial fluctuations, $\delta T$ and $\delta
\omega$, to reach their steady state values are diffusive, and increase
as $k_{\rm min}^{-2}\sim L^2$ with system size.
We have checked in the simulations
that the large scale behavior of the structure factors
was properly equilibrated before accumulating the data
used to compute the averages.

First of all, we have tested the isotropy of tensor (\ref{eq:Smomentum})
by checking that the average
\be
\biggl\langle \sum_{i,j} \left(
\bv_i\cdot\widehat{\bk}\right) \left( \bv_j\cdot\widehat{\bk}_{\perp}
\right)\,
\exp(\imath\, \bk\!\cdot\!\br_{ij})
\biggr\rangle,
\qquad \hbox{with}\quad \widehat{\bk}\cdot\widehat{\bk}_\perp = 0,
\ee
vanishes for $\bk$ values compatible with the periodic boundary
conditions.

In Fig.\ \ref{fig:S0.92} we show the density-density and the relevant
components of the velocity-velocity structure
factors.
For {\em elastic} hard disks the plateau values of $S_{aa}(k)$ around
$k\sigma \simeq 2$ extend all the way down to $k=0$. The excess
correlations in the dissipative regime ($k \lesssim \gamma_0/l_0$),
which for
$S_\perp(k)$ extend up to  
$k \simeq \sqrt{\gamma_0}/l_0$, are characteristic of the randomly
driven inelastic fluid.
Fig.\ \ref{fig:S0.92} shows that for
small inelasticities
the agreement between
simulations and theory
is quite reasonable.
The structure factors diverge at small
scales like $k^{-2}$, in agreement with the theoretical predictions
(\ref{eq:sperp})-(\ref{eq:bleue}).
The packing
fraction has been chosen fairly high ($\phi=0.63$) but lower than
the two-dimensional random close-packing of monodisperse disks
$\phi_{RCP}\simeq 0.82$
\cite{Jasty},
and inside the {\em liquid} region of the phase diagram for 
elastic hard disks.
In addition to the gain in computer time, such a
packing fraction leads to a mean free path $l_0$ smaller than the
particle diameter $\sigma$ (e.g.\ $l_0\simeq 0.095\sigma$ for $\phi=0.63$).
Therefore, the hydrodynamic regime will hold up to
typical particle diameters, enlarging the range of wave vectors where
comparison between simulations and theoretical predictions is
feasible. Moreover, for dense systems, a marked
density structure is to be expected at the molecular
scale, especially at wavelengths close
to $\sigma$ ($k\sigma \simeq 2\pi$). Fig.\ \ref{fig:S0.92_ehd}
shows that for $\alpha$ close to 1,
this structure is indistinguishable from the
structure for elastic hard disks.
Note that $S_{ll}(k)$ and $S_{\perp}(k)$,
although quite structureless for $k\sigma>1$, show a weak
and broad peak correlated with the maximum of $S_{nn}(k)$.
This feature is more pronounced as the inelasticity
increases, as shown in Figs.\ \ref{fig:S0.6tout} and \ref{fig:S0tout}.
As the molecular structure of the fluid has not been taken into
account in the long wavelength hydrodynamic approach of section \ref{sec:strf},
the present theory cannot explain the structure in Figs.\ \ref{fig:S0.6tout} 
and \ref{fig:S0tout}, and
the structure factors predicted by the theory reach a 
plateau in the elastic regime ($k l_0 \gtrsim \sqrt{\gamma_0}$), given by
\begin{eqnarray}
&&S_{nn}(k) \to n^2 \,  T\, \chi_{_{T}} \nonumber \\
&&S_{\alpha\beta}(\bk) \to \frac{T}{\rho} \,\delta_{\alpha\beta},
\end{eqnarray}
where $\chi_{_{T}}$ is the isothermal compressibility of the elastic hard
disk fluid. For $\alpha$ close to 1,
the above limiting behavior is observed numerically
for $S_{ll}$ or $S_{\perp}$, and for $S_{nn}$ only in the
limit of small packing fraction, where the molecular
structure disappears.

When the inelasticity is increased, the structure factors exhibit the same
$k^{-2}$ behavior at large scale, but the theoretical
expressions are less accurate (see Fig.\ \ref{fig:S0.6hydro}).
However, the theoretical curves
are based on the Enskog estimate $T_{\rm E}$ for the
granular temperature $T$, whereas for the system
corresponding to Fig.\ \ref{fig:S0.6hydro}, $T \simeq 1.45
T_{\rm E}$.
Our mode coupling theory predicts here $T_{\rm NESS}\simeq
1.41 T_{\rm E}$.
Fig.\
\ref{fig:S0.6hydro1.45} displays the comparison between theory
and simulation when the measured granular temperature
is taken as an input for the hydrodynamic description.
It appears that the large scale correlations (for which the present
theory has been constructed) are well described
by the theory, as long as the temperature
is corrected from the mean field Enskog prediction (\ref{eq:Tss})
to the measured value $T$. In the case of $S_{\perp}$, the
amplitude only depends on the shear viscosity. The good agreement of
the amplitude when the temperature is rescaled, while keeping
for the shear viscosity the elastic hard
disk value, suggests
that the dependence of the shear viscosity on the inelasticity
could be attributed only
to the change in temperature.
At the molecular scale, $S_{ll}(k)$ and $S_{\perp}(k)$
appear to be correlated to the density-density structure
factor (see Fig.\ \ref{fig:S0.6tout}), in marked contrast
to the elastic situation, where a plateau value would be reached.
Such an effect is beyond the scope of our hydrodynamic
approach, and is currently under investigation.

Surprisingly, in the case of complete inelasticity ($\alpha=0$),
the theoretical structure factors give a reasonable picture
of the large wavelengths correlations in the fluid
(especially for the longitudinal velocity component $S_{ll}(k)$, as shown
in Fig.\ \ref{fig:S0hydro0.76}). When the theoretical structure factors
are deduced
from the measured granular temperature $T\simeq 1.5 T_{\rm E}$ 
and not from $T_{\rm E}$ (our mode coupling theory predicts $T_{\rm
NESS}\simeq 1.77 T_{\rm E}$),
the agreement
for $S_\perp(k)$ improves, but the mismatch for $S_{ll}(k)$
increases (see Fig.\ \ref{fig:S0hydro}). At small scales,
the density correlations differ significantly from
the elastic ones (Fig.\ \ref{fig:S0tout}) and the velocity structure factors
exhibit the oscillatory behavior already present in Fig.\
\ref{fig:S0.6tout}, with peak positions locked in on the
peaks in $S_{nn}(k)$. As can be expected from Figs.\ \ref{fig:S0hydro0.76} and
\ref{fig:S0hydro}, the large scale correlations are compatible
with the expected $k^{-2}$ law (see Fig.\ \ref{fig:S0log})
unlike the results of Peng and Ohta who report a
$k^{-1.4}$ asymptotic behavior for $\alpha=0$.
However, a log-log plot such as Fig.\ \ref{fig:S0log}
does not allow an accurate evaluation of the scaling
exponents. The $k^{-2}$ law is better inferred
from the direct comparison with theory (Figs.\
\ref{fig:S0.92}, \ref{fig:S0.6hydro1.45}, \ref{fig:S0hydro}).

Before concluding this section we compare the structure factors for the
longitudinal flow fields, obtained from MD simulations with two
different theoretical predictions in Fig.\ \ref{fig:int-ext}, obtained
by including or excluding internal noise.
First observe that all parameters in Fig.\ \ref{fig:int-ext} and Fig.\
\ref{fig:S0.92} are identical, as well as units on both axes.
The simulation results for $S_\parallel(k)$ in Fig.\ \ref{fig:S0.92}
are in excellent agreement with the 
theoretical prediction (dashed line), which corresponds to
the solid line in Fig.\ \ref{fig:int-ext} (internal plus external
noise).
The dotted line (without internal noise)
for $S_\parallel^{\rm without}(k)$ in Fig.\
\ref{fig:int-ext} disagrees with the simulations in the relevant
interval $0.2 \lesssim k\sigma \lesssim 0.5$.

Hence, inclusion of internal noise extends the validity of the
asymptotic theory to intermediate wave numbers.

\section{Conclusion}
\label{sec:conc}
We have presented a theory for the large scale dynamics of a granular
fluid that is driven into a nonequilibrium steady state
by a random external force.
Our description combines the macroscopic equations of motion for the
hydrodynamic fields, accounting for energy dissipation through
inelastic collisions and uniform heating, together with the fluctuating
forces.
The long range character of the spatial correlation functions is
determined by the small wave number divergence $\sim k^{-2}$ of the
corresponding structure factors.
This $k^{-2}$ behavior is typical for systems which combine conserving
deterministic dynamics (conservation of particle number and momentum in
collisions) with nonconserving noise \cite{grinstein}, thus violating
the fluctuation-dissipation condition, and is generic
for rapid granular flows that are
driven by external noise.
We also draw attention to the analogy of our equations of motion for
the fluctuating fields to the Edwards-Wilkinson model \cite{edwards}
that was proposed for growth of a granular surface.
In that case the dynamic variable is a scalar field, namely the height of
the surface, which obeys a similar equation of motion as 
any of the $(d-1)$ components of the transverse velocity field,
$u_{\perp\alpha}(\bk,t)$, in our case.
The only difference is that in the Edwards-Wilkinson model there is only
nonconserving noise, whereas in our case both
nonconserving (external) and conserving (internal) noise are present.

We have tested our predictions for the structure factors
against molecular dynamics simulations and
have demonstrated that there is quantitative agreement over the whole
wave number range, if internal fluctuations are taken into account.
In our two-dimensional simulations
we have found
deviations in the steady state temperature and collision frequency
that grow with the inelasticity and the system size.
For not too large inelasticities ($\alpha\gtrsim 0.6$),
we have explained these deviations in terms of mode coupling effects of
the long range fluctuations.

The phenomenological mode coupling theory, proposed in \cite{brito+ernst}
and extended here to the driven 
IHS fluid, 
starts from the same basic ingredients as in the case
of elastic fluids \cite{EHvL}, where that theory was used to
calculate the long time tails of the velocity autocorrelation function
and other current-current time correlation functions.
For the elastic case the mode coupling
theory can be derived from the {\em ring
kinetic theory} in the low density limit \cite{dced}, which accounts for
dynamic correlations built up by sequences of correlated binary
collisions, leading to nonlocal effects in space and time.
Such collision sequences correct the mean field-type Boltzmann or
Enskog kinetic equations for the errors induced by the breakdown of
the molecular chaos assumption.
The ring kinetic theory for rapid granular flows of IHS has been
developed in Ref.\ \cite{vleeuwen}, but has not yet been used to
derive the present phenomenological mode coupling theory from the more
fundamental kinetic theory, valid in the low density limit.

In detailed balance models, such as elastic hard spheres, dynamic
correlations created by correlated collision sequences lead to long
time tails, which imply that 
transport
coefficients in two dimensions diverge as $\ln{L}$ for large
systems \cite{EHvL,dced}.

Non-detailed balance models, such as IHS fluids, generically exhibit
long range spatial correlations \cite{dorfman}.
In randomly driven IHS fluids, as studied in this paper, 
these correlations between densities and
flow fields at distant points in the fluid behave as $1/r$ in 3D and
$\ln{r}$ in 2D, and already
modify (renormalize) the mean field Enskog-Boltzmann values
for steady state properties, such as the temperature and collision
frequency.
In 2D systems these renormalization corrections, $\delta T$ and
$\delta \omega$, exhibit the $\ln{L}$ divergence, which in the
case of detailed balance models appears only in the transport
coefficients \cite{dorfman}.

At larger inelasticity ($\alpha \lesssim 0.6$), molecular chaos is
also violated due to the presence of short range
velocity-velocity correlations
(see Fig.\ \ref{fig:S0tout}), which are beyond our mode coupling
theory.
A detailed investigation of the small scale structure, which for large
inelasticity clearly deviates from an equilibrium structure, will be
reported in a subsequent publication.
It is surprising that our description, which is based on the Enskog theory
and neglects any dependence of transport coefficients on inelasticity,
even at $\alpha=0$ predicts the long range structure reasonably well,
provided that the temperature is not taken as the mean field Enskog
value, but set equal to the value measured in the simulations.

\section*{Acknowledgments}
The authors acknowledge useful discussions with
D. Frenkel and I. Goldhirsch. E.T. and I.P. thank
D. Frenkel and B. Mulder for the hospitality of their groups at AMOLF.
T.v.N. and I.P. acknowledge support of the
foundation `Fundamenteel Onderzoek der Materie (FOM)', which is
financially supported by the Dutch National Science Foundation
(NWO). 

\newpage
{\appendix
\section{}
\label{sec:appA}
In this appendix we derive expressions for the transport coefficients that
govern the
decay of fluctuations in the steady state.
Linearization of the macroscopic equations (\ref{eq:hydro}) around the
NESS, defined in (\ref{eq:Tdec}) and (\ref{eq:Tss}), gives the deterministic
part of the following set of
equations:
\ba
\partial_t \delta n &=& - n \bnabla\cdot\bbox{u}\nonumber\\
\partial_t \bbox{u} &=& - \frac{1}{\rho}\bnabla p
+\nu \nabla^2 \bbox{u}+(\nu_l-\nu) \bnabla \bnabla\cdot\bbox{u}+
\hat{\bxi}\nonumber\\
\partial_t \delta T &=& \frac{2\kappa}{dn}\nabla^2\delta T -\frac{2 p}{d
n}\bnabla\cdot\bbox{u}-\delta \Gamma+\hat{\theta}.
\label{eq:app1}
\ea
The noise terms $\hat{\bxi}(\br,t)$ and $\hat{\theta}(\br,t)$ have been
discussed in section \ref{sec:noise}.
The pressure $p$ is assumed to be that of EHS, $p=n T(1+\Omega_d\chi n
\sigma^d /2 d)$, where $\Omega_d=2 \pi^{d/2}/\Gamma(d/2)$ is the
$d$-dimensional solid angle, and $\chi(n)$ the equilibrium value of
the pair correlation function of EHS of diameter $\sigma$ and mass $m$ at
contact.
The kinematic and longitudinal viscosities $\nu$ and $\nu_l$, as well
as the heat conductivity $\kappa$ are also assumed to be approximately
equal to the corresponding quantities for EHS, as calculated from the
Enskog theory \cite{chapman}, where $\rho\nu=\eta$ and $\rho
\nu_l=2\eta(d-1)/d+\zeta$ are expressed in shear viscosity $\eta$ and
bulk viscosity $\zeta$.
The collisional energy loss in (\ref{eq:Gloss}),
$\Gamma=2\gamma_0\omega T$, is proportional to the collision frequency
\be
\omega =\Omega_d \chi n \sigma^{d-1}\sqrt{\frac{T}{\pi m}},
\label{eq:app2}
\ee
as obtained from the Enskog theory.
In two dimensions we use the Verlet-Levesque approximation $\chi=
(1-7\phi/16)/(1-\phi)^2$.

By taking spatial Fourier transforms $\delta a(\bk,t) = \int {\rm
d}\br \delta a(\br,t) \exp(-\imath \bk\cdot\br)$ in (\ref{eq:app1}),
one obtains the
mesoscopic equation (\ref{eq:meso}) with the hydrodynamic matrix
\ba
{\bf M}(\bk)=-\left( \begin{array}{cccc}
0 & 0 & i k n & 0\\
\gamma_0 \omega g(n)T/n & 3 \gamma_0 \omega+D_T k^2 & i k 2 p/d n&0\\
i k v_T^2/n & i k p/\rho T & \nu_l k^2 & 0\\
0 & 0& 0& \nu k^2
\label{eq:app3}
\end{array}\right).
\ea
It contains the coefficients
\ba
g(n)&=&2\left(1+\frac{n}{\chi}\frac{{\rm d}\chi}{{\rm d}n}\right)\nonumber\\
v_T^2&=&\left(\frac{\partial p}{\partial
\rho}\right)_T\nonumber\\
D_T&=&\frac{2\kappa}{d n}.
\label{eq:app4}
\ea
In the body of the paper we need the eigenvalues $z_\lambda(\bk)$
of the asymmetric matrix $M_{ab}(\bk)$, and its
right and left eigenvectors, which are obtained from
\ba
{\bf M}(\bk)\bw_\lambda(\bk)&=&z_\lambda(\bk)
\bw_\lambda(\bk)\nonumber\\
{\bf M}^T(\bk)\bv_\lambda(\bk)&=&z_\lambda(\bk) \bv_\lambda(\bk),
\label{eq:ev}
\ea
where ${\bf M}^T$ is the transpose of ${\bf M}$.
Here
$\lambda=\pm$
labels the sound modes, $\lambda=H$ the heat mode, and $\lambda=\perp$
labels
$(d-1)$ degenerate shear or transverse velocity modes.
The eigenvectors form a complete biorthonormal basis, i.e.\
\ba
\sum_a v_{\lambda a}(\bk) w_{\mu a}(\bk)&\equiv&\langle \bv_\lambda|
\bw_\mu\rangle=\delta_{\lambda\mu}\nonumber\\
\sum_\lambda |\bw_\lambda(\bk)\rangle  \langle
\bv_\lambda(\bk)|&=&\bbox{I}.
\label{eq:compl}
\ea
Moreover the eigenvalue equation, ${\rm det}[z(k)\bbox{I}-{\bf
M}(\bk)]=0$, is an {\em even} function of $\bk$.
Consequently, ${\bf M}(\bk)$ and ${\bf M}(-\bk)$ have the same
eigenvalues, which are either real or form a complex conjugate pair.
So we choose $z_\lambda(\bk)=z_\lambda(-\bk)=
z_\lambda(k)$.
The corresponding eigenvectors of ${\bf M}(-\bk)$ in case of the sound
modes are obtained from the
transformation $\{\bw_+(-\bk),\bv_-(-\bk)\}\leftrightarrow
\{\bw_{-}(\bk),\bv_{+}(\bk)\}$.
All other eigenvectors are invariant under the transformation
$\bk\rightarrow -\bk$.

By setting $z(k)=0$ in the eigenvalue equation,
one can verify that there are no
zero crossings at any finite wave number.
So, all eigenvalues have a definite (here {\em negative}) sign.
Consequently all modes of the heated IHS fluid are linearly stable.
There is {\em no} clustering instability \cite{noije,goldhirsch}
and {\em no} instability in the flow field \cite{noije},
as in the freely evolving
IHS fluid.

In the {\em dissipative regime} ($k l_0\ll \gamma_0$) the eigenvalue
equation is solved by an expansion in powers of $k$, and one finds to
dominant orders the eigenvalues in the form (\ref{eq:eigenva}) and
eigenvectors in the form (\ref{eq:eigenve}).
The eigenmodes to dominant nonvanishing order in $k$ are listed in
(\ref{eq:eigenve}).
There are $(d-1)$ transverse velocity or shear modes ($\lambda=\perp$),
which are purely diffusive with a diffusivity $D_\perp=\nu$;
there are two propagating modes ($\lambda=\pm$)
with a speed of propagation $v_D$, and
sound damping constant $D_S$, and a {\em
kinetic} heat mode ($\lambda=H$) with a nonvanishing $z_H(0)$.

For later reference we also express these coefficients in
thermodynamic quantities and transport coefficients.
The speed of sound $v_D$ in the {\em dissipative} regime ($k l_0\ll
\gamma_0$) is:
\be
v_D^2=\left(\frac{\partial p}{\partial
\rho}\right)_T-\frac{2 p}{3\rho}\left(1+\frac{n}{\chi}\frac{{\rm
d}\chi}{{\rm d}n}\right).
\label{eq:vD}
\ee
It satisfies the inequality $v_D<v_S$, where $v_S$ is the
adiabatic speed of sound in the {\em standard} regime ($\gamma_0\ll
k l_0<1$),
\be
v_S^2=v_T^2+ \left(\frac{2 p}{d n}\right)
\left(\frac{p}{\rho T}\right)=\left(\frac{\partial
p}{\partial
\rho}\right)_S.
\ee
The damping constant of the sound modes is
\be
D_S={\textstyle{\frac{1}{2}}} \nu_l +
\frac{p}{9\gamma_0\omega\rho}\left(1+\frac{n}{\chi}\frac{{\rm
d}\chi}{{\rm d}n}+\frac{3 p}{dnT}\right),
\label{eq:Ds}
\ee
and the dispersion relation for the kinetic heat mode contains the
positive constant
\be
D_H= -\frac{2\kappa}{dn} +\frac{2p}{9\gamma_0
\omega \rho}\left(1+\frac{n}{\chi}\frac{{\rm
d}\chi}{{\rm d}n}+\frac{3 p}{dnT}\right).
\label{eq:DH}
\ee
The ratios $v_D^2/T$, $v_S^2/T$, $D_S/\sqrt{T}$, and $D_H/\sqrt{T}$
are independent of temperature.

\section{}
\label{sec:appB}
In this appendix we calculate the tails of
the spatial correlation functions $G_{ab}(\br)$, which
is done by Fourier
inversion of $S_{ab}(\bk)$.
Consider first the tensor fields $G_{\alpha\beta}(\br)$ and
$S_{\alpha\beta}(\bk)$ in Eq.\ (\ref{eq:strfa}) with $\delta a_\alpha(\br,t)=
u_\alpha(\br,t)$ ($\alpha,\beta=x,y,\dots$) being the components of the
flow field.
Both tensor fields are isotropic and can be split into longitudinal and
transverse components, i.e.\
\ba
G_{\alpha\beta}(\br)&=&\hat{r}_\alpha\hat{r}_\beta
G_\parallel(r)+(\delta_{\alpha\beta}-\hat{r}_\alpha\hat{r}_\beta)
G_\perp(r)\nonumber\\
&=&\int\frac{{\rm d}\bk}{(2\pi)^d} \exp(\imath \bk\cdot \br)\left[
\hat{k}_\alpha \hat{k}_\beta S_\parallel(k) + (\delta_{\alpha\beta}
-\hat{k}_\alpha\hat{k}_\beta)S_\perp(k)\right],
\label{eq:appb1}
\ea
where the {\em small} $k$ behavior of $S_\perp(k)$ and $S_\parallel(k)$ is
given in (\ref{eq:sperp}) and (\ref{eq:sparallel}).
By contracting the second line above with $\hat{r}_\alpha\hat{r}_\beta$
we obtain
\be
G_\parallel(r)=\int\frac{{\rm d}\bk}{(2\pi)^d} \exp(\imath \bk\cdot
\br) \left[(\hat{\bk}\cdot\hat{\br})^2 S_\parallel(k)
+[1-((\hat{\bk}\cdot\hat{\br})^2]S_\perp(k)\right].
\label{eq:appb2}
\ee
Contraction of the second line of (\ref{eq:appb1}) with
$(\delta_{\alpha\beta}
-\hat{k}_\alpha\hat{k}_\beta)$ yields in a similar manner an expression
for $G_\perp(r)$.

In {\em three} dimensions the $\bk$ integral can be performed
explicitly and yields
\ba
\int \frac{{\rm d}\bk}{(2\pi)^3} \frac{\exp(\imath \bk\cdot
\br)}{k^2} &=& \frac{1}{4\pi r}\nonumber\\
\int \frac{{\rm d}\bk}{(2\pi)^3} \frac{\exp(\imath \bk\cdot
\br)}{k^2} (\hat{\bk}\cdot\hat{\br})^2 &=& 0.
\ea
The resulting large $r$ behavior of $G_{\alpha\beta}(\br)$ is given in
(\ref{eq:large3}).

In {\em two} dimensions the integral in (\ref{eq:appb2}) over the
azimuthal angle yields for large $r$
\ba
\int \frac{{\rm d}\bk}{(2\pi)^2} \frac{\exp(\imath \bk\cdot
\br)}{k^2} &=&
\frac{1}{2\pi}\int_{k_{\rm min}}^\infty \frac{{\rm d}k}{k} J_0(kr)
\simeq\frac{1}{2\pi} \ln\left(\frac{L}{r}\right)+{\cal
O}(1)\nonumber\\
\int \frac{{\rm d}\bk}{(2\pi)^2} \frac{\exp(\imath \bk\cdot
\br)}{k^2} [1-(\hat{\bk}\cdot\hat{\br})^2]&=&\frac{1}{2\pi
r}\int_{k_{\rm min}}^\infty \frac{{\rm d}k}{k^2}J_1(kr)\simeq
\frac{1}{4\pi} \ln\left(\frac{L}{r}\right)+{\cal
O}(1).
\ea
In two dimensions the $\bk$ integral diverges for $k\rightarrow 0$.
It should in fact be restricted to $k\geq k_{\rm min} = 2 \pi/L$, which is
the smallest allowed wave number when periodic boundary conditions are
imposed.
To obtain the last equalities one needs the small $z$ behavior of the
Bessel functions $J_\nu(z)\simeq (z/2)^\nu /\Gamma(\nu+1)$
\cite{gradshteyn}.
Combining these results with (\ref{eq:appb2}) yields for large $r$
\be
G_\parallel(r)\simeq G_\perp(r)\simeq\frac{\Gamma}{8\pi \rho}
\left(\frac{1}{\nu}+\frac{1}{2 D_S}\right)
\ln\left(\frac{L}{r}\right).
\ee
The remaining spatial correlation functions $G_{ab}(r)$ involving
$a,b=\{n,T\}$, are scalar fields.
Their large $r$ behavior is given by
\be
G_{ab}(r)=\int\frac{{\rm d}\bk}{(2\pi)^d} \exp(\imath \bk\cdot
\br) S_{ab}(k)= \frac{\Gamma B_{ab}}{8 \pi \rho D_S}\left\{ \begin{array}{ll}
\ln(L/r) \,\,\,\,\,(d=2)\\
1/2r \,\,\,\,\,\,\,\,\,\,\,\,(d=3),
\end{array}\right.
\ee
where $B_{ab}(k)$ is given in Table \ref{bleue}.
The subleading corrections of ${\cal O}(r^0)$ depend on the cutoff
$k_{\rm min}$.
To evaluate these terms requires the subleading small $k$ behavior of
$S_{ab}(k)$ in Eqs.\ (\ref{eq:sperp}) to (\ref{eq:bleue}).

\section{}
\label{sec:appC}
In this appendix we present a mode coupling calculation to estimate the
contributions of the long wavelength fluctuations in the NESS
to some quantity $h$.
Examples are the particle distribution functions, the energy per particle
$E$, and the collision frequency $\omega$.
The fluctuations are correlated over large distances, as a consequence
of sequences over dynamically correlated collisions, the so called ring
collisions \cite{berne}, as shown in section \ref{sec:analytical}
by explicit calculation of their spatial tails.

To calculate their contributions to average quantities like $h$, one may
solve the ring kinetic equations \cite{vleeuwen}, or estimate
these quantities from a more phenomenological mode coupling approach, as
developed in Ref.\ \cite{EHvL}.
The basic assumption made there is that the state of the system rapidly
decays to a state of local equilibrium, described by the fluctuating
hydrodynamic fields ${\bf a}(\br,t) = \{n(\br,t), u_\alpha(\br,t),
T(\br,t)\}$.

For the quantity under consideration this can be implemented by
representing $h=(1/V)\int {\rm d}\br h(\br)$, and approximating $h(\br)$
by its value in local equilibrium, i.e.\
\be
h_{\rm NESS}=\frac{1}{V}\int {\rm d}\br \langle h_l\left({\bf a}(\br)
\right)\rangle,
\label{eq:appc1}
\ee
where the average is taken over the fluctuating hydrodynamic fields ${\bf
a}(\br)$ in the NESS.
To carry out the average over the fluctuations we expand $h_l({\bf a})$
in powers of the fluctuations $\delta a = a-\langle a\rangle$ around the
NESS, yielding
\ba
h_{\rm NESS}&\simeq& h_l(\langle {\bf a}\rangle) + \frac{1}{2V}\int^\ast {\rm
d}\br \langle
\delta a(\br) \delta b(\br)\rangle A_{ab}\nonumber\\
&=& h_l(\langle {\bf a}\rangle) + \frac{1}{2} \int^\ast\frac{{\rm
d}\bk}{(2\pi)^d} S_{ab}(k)
A_{ab},
\label{eq:hness}
\ea
where summation convention for repeated indices has been used.
Here $A_{ab}$ is the
matrix of second derivatives $\partial^2 h_l({\bf a})/\partial {\bf a}
\partial {\bf a}$ at
${\bf a}=\langle {\bf a}\rangle$.
The asterisk indicates that $\bk$ integrals are restricted to the long
wavelength range, $k<\gamma_0/l_0$, the so called dissipative range,
discussed below (\ref{eq:ineq}).
In this range the structure factors have the form $S_{ab}(k)\simeq
E_{ab}/k^2$ on account of (\ref{eq:sperp}) to (\ref{eq:bleue}).

For dimensionality $d \ge 3$ the fluctuation contribution, $\delta
h=h_{\rm NESS}-h_l(\langle {\bf a}\rangle)$, is convergent at small $k$,
and gives only small
well-behaved corrections to $h_l(\langle {\bf a}\rangle)$.
However, for $d=2$, the $\bk$ integral diverges logarithmically at small
$k$ (where $k\gtrsim 2\pi/L$), and the excess, $\delta h$, is given by
\be
\delta h\simeq \frac{A_{ab}E_{ab}}{4\pi} \int_{2\pi/L}^{\gamma_0/l_0}
\frac{{\rm d}k}{k}
\simeq \frac{A_{ab}E_{ab}}{4\pi}\ln\left( \frac{\gamma_0 L}{l_0}\right).
\ee
Consequently, the fluctuation contribution $\delta h$ in 2D systems
is a singular
function of the system size, $L$, that diverges in the thermodynamic
limit.

We first apply the above results to the energy per particle,
$E=(1/N)\int {\rm d}\br
e_l\left({\bf a}(\br)\right)$, where
$e_l({\bf a})=\textstyle{\frac{1}{2}} \rho
u^2 + \textstyle{\frac{d}{2}} n T$ is the energy density in local
equilibrium.
From $e_l({\bf a})$ the expansion coefficients corresponding to
$h_l(\langle{\bf a}\rangle)$
and $A_{ab}$ in (\ref{eq:hness}) can be calculated, to yield in $d=2$:
\be
\delta E\simeq \frac{1}{2 n} \int^\ast\frac{{\rm d}\bk}{(2\pi)^2} \left[\rho
S_\perp(k)+\rho S_{ll}(k)+ 2 S_{nT}(k)\right].
\ee
Inserting (\ref{eq:sperp}) to (\ref{eq:bleue}) then yields
\be
\delta E \simeq \frac{\gamma_0 \omega_{\rm E} T_{\rm E}}{4\pi n}
\left[\frac{1}{\nu}+\frac{1}{ 2 D_S}
\left(1+\frac{2 B_{nT}}{\rho}\right)\right] \ln\left( \frac{\gamma_0
L}{l_0}\right),
\ee
where the coefficient
$B_{nT}$ is listed in Table \ref{bleue}.

For the collision frequency the analog of (\ref{eq:appc1}) is
$\omega=(1/N) \int {\rm d}\br \langle n(\br) \omega_l\left({\bf
a}(\br)\right)\rangle$ with $\omega_l({\bf a})\propto n \chi(n)
\sqrt{T}$ given in (\ref{eq:app2}).
This gives in two dimensions:
\ba
\delta \omega &\simeq& \frac{1}{2 n} \int^\ast\frac{{\rm
d}\bk}{(2\pi)^2}\left[ S_{nn}(k) A_{nn} +2 S_{nT}(k) A_{nT} + S_{TT}(k)
A_{TT}\right]\nonumber\\
&\simeq& \frac{\gamma_0 \omega_{\rm E} T_{\rm E}}
{8 \pi n \rho D_S}\left[ B_{nn} A_{nn} +2 B_{nT}
A_{nT} + B_{TT} A_{TT}\right] \ln\left( \frac{\gamma_0
L}{l_0}\right),
\ea
where the coefficients $B_{ab}$ are given in Table \ref{bleue}, and
$A_{nn}=\partial^2 (n\omega)/\partial n^2$, $A_{nT}=\partial^2
(n\omega)/\partial n\partial T$, and $A_{TT}=\partial^2
(n\omega)/\partial T^2$ at ${\bf a}=\langle {\bf a}\rangle$.

The same method can be used to calculate other averages, as well as
particle distribution functions.
For instance, for the single particle distribution function the
starting point would be
\be
f_{\rm NESS}(\bv)=(1/V)\int {\rm d}\br \langle f_l\left(\bv|{\bf
a}(\br)\right)\rangle,
\ee
and the above procedure can be applied at once.


\begin{table}
\begin{tabular}{|c|c|}
$ab$&$B_{ab}$\\ \hline
$ll$&$1$\\
$nn$&$n^2/v_D^2$\\
$TT$&$g^2(n) T^2/ 9 v_D^2$ \\
$nT$& $-g(n) n T/ 3 v_D^2$ \\
\end{tabular}
\caption{Coefficients $B_{ab}$ in Eq.\ (\ref{eq:bleue}).}
\label{bleue}
\end{table}

\begin{table}
\begin{tabular}{|c|c|c|c|c|c|}
$\Theta$   & 1.5$^o$ & 3.5$^o$ & 5$^o$  & 45$^o$  & 90$^o$ \\ \hline
~$\omega/\omega_{\rm E}$~ & 8.2   & 7.4 & 7  & 4.3  & 4.3 \\
\end{tabular}
\caption{Collision frequency (normalised by the Enskog value)
as a function of $\Theta$, for a totally inelastic system ($\alpha=0$)
with $N=1600$ particles, and packing fraction $\phi = 0.07$.}
\label{rouge}
\end{table}

\begin{center}
{FIGURE CAPTIONS}
\end{center}

\begin{enumerate}
\item Dispersion relations $z_\lambda(k)/\omega$
versus $k\sigma$ for $\phi=0.4$,
$\alpha=0.9$; the solid lines refer to the real parts for
$\lambda=\perp,\pm$, and $H$ respectively. Dashed lines represent the
imaginary parts of the sound mode relaxation rates ($\lambda=\pm$).
Here, $l_0/\sigma\simeq 0.34$, $\gamma_0\sigma/l_0\simeq 0.14$, and
$\sqrt{\gamma_0}\sigma/l_0\simeq 0.64$.

\item
Structure factors, $S_\perp(k)$ and $S_\parallel(k)$, as obtained from
the full theory (solid lines) with external {\em and} internal noise.
The dotted line represents $S_\parallel^{\rm without}(k)$ without
internal noise with the plateau value added (see discussion at the end
of section \ref{sec:simul}).
The parameters are $\alpha=0.92$ and $\phi=0.63$.
Fig.\ \ref{fig:S0.92} shows that the $S_\parallel$ simulation data
agree much better with the solid line than with the dotted line.

\item Granular
temperature as a function of time for $\alpha=0.92$, $\phi = 0.078$
($l_0/\sigma\simeq 3.5$),
and $N=1600$ particles. The simulation result is compared
to the analytical expression (\ref{eq:Tdet}) (dashed curve). The initial
condition corresponds to a fluid-like configuration of elastic hard
disks.
Here $\Delta t = 3\,10^{-5}\, (mL^2/T_0)^{1/2}\simeq 1.5\, 10^{-3}\,
t_0$
and $\varphi_0 = 5.77\, 10^{-2}\, (T_0/m)^{1/2}$. $T_{\rm E}$
is the temperature
expected on the basis of the Enskog theory (see Eq.\ (\ref{eq:Tss})). For the
above parameters, there are on average 3.7 collisions per time interval
$\Delta t$ in the NESS, and $T_{\rm E}/T_0=9.3$.

\item Time dependence of the granular
temperature obtained in the simulation
for the two-dimensional
system of Fig.\ \ref{fig:Tdet}, with $\Delta t = 3\, 10^{-4}\, 
(mL^2/T_0)^{1/2}$,
$\varphi_0 = 1.7\, 10^{-3} (T_0/m)^{1/2}$. Comparison is made
with Haff's law (\ref{eq:homco}) for
homogeneous cooling
(dashed curve), and with the full
solution of Eq.\ (\ref{eq:Tdet}) (long-dashed curve).
Here, $T_{\rm E}/T_0 = 0.018$.

\item Measured excess temperature, $\delta T=T-T_{\rm E}$, and
collision frequency, $\delta \omega=\omega -\omega_{\rm E}$, versus
coefficient of restitution $\alpha$, for $L=60\sigma$, $N=917$
($\phi=0.2$, $l_0=1.1\sigma$), and for 
two different values of the maximal random rotation angle,
$\Theta=5^o$ and $\Theta = 10^o$.
Inset: comparison with
the predictions of the mode coupling theory, Eq.\ (\ref{eq:eoness}).

\item Structure factors
versus wave vector for $\alpha=0.92$, $\phi = 0.63$
($l_0/\sigma\simeq 0.095$), and $N=10201$. The simulation
data (symbols) have been averaged over $10^2$ successive configurations,
separated by a time interval of 20 collisions per particle. $S_{ll}$ and
$S_{\hbox{\scriptsize \it perp}} \equiv S_{\perp}$ are respectively the
parallel and perpendicular parts of the velocity-velocity
structure factor, defined by Eq.\ (\ref{eq:Smomentum}).
Comparison is made with the theoretical expressions (full, dashed and
long-dashed curves)
deduced from Eq.\ (\ref{eq:staticsf}) (compare also Fig.\ \ref{fig:int-ext}).
There is a dissipative regime for $k\sigma \lesssim
\gamma_0 \sigma/l_0 \simeq 0.40$, and an elastic regime for $k\sigma \gtrsim
\sqrt{\gamma_0}\sigma/l_0 \simeq 2.1$.
Here, $T/T_{\rm E}\simeq \omega/\omega_{\rm E} \simeq 1.05$
whereas  the mode coupling approach of section
\ref{sec:analytical}
predicts $T_{\rm NESS}/T_{\rm E}\simeq 1.07$ and
$\omega_{\rm NESS}/\omega_{\rm E} \simeq 1.003$.

\item Same as Fig.\ \ref{fig:S0.92}, with $k$ beyond the dissipative regime.
The dashed curve corresponds to the density-density structure factor
of an elastic hard disk (EHD) system of the same size and packing
fraction (dashed curve). All results are deduced from
MD simulations.

\item Structure factors for $\alpha = 0.6$, $\phi=0.55$
($l_0/\sigma\simeq 0.15$), and $N=10201$.
The lines are the corresponding theoretical predictions.
The measured temperature $T/T_{\rm E}\simeq 1.45$ and our mode
coupling theory gives $T_{\rm NESS}/T_{\rm E}\simeq 1.41$.

\item Same as Fig.\ \ref{fig:S0.6hydro} where for the theoretical
expressions (lines),
the temperature has been set to the measured
kinetic energy per particle, $T\simeq 1.45 T_{\rm E}$.

\item Same as Fig.\ \ref{fig:S0.6hydro} beyond the dissipative regime.
The theoretical structure factors are not displayed.

\item Structure factors for $\alpha=0$, $\phi=0.63$ and $N=10201$.
Comparison is made with the hydrodynamic theory (lines).
Here, $T/T_{\rm E}\simeq 1.5$, whereas mode coupling gives $T_{\rm
NESS}/T_{\rm E}\simeq 1.77$.

\item Same as Fig.\ \ref{fig:S0hydro0.76}, where the simulation data
(symbols) are compared to the theoretical predictions
(lines) for which the temperature has been set to the measured
kinetic energy per particle, $T\simeq 1.5 T_{\rm E}$.

\item Structure factors up to the molecular scale, for the same
parameters as in Fig.\ \ref{fig:S0hydro0.76}.
$S_{nn}$ for elastic hard disks and the same packing fraction
has also been plotted (crosses).

\item $\log_{10}|S-S_{min}|$ versus $\log_{10}(k\sigma)$, for
the same parameters as in Fig.\ \ref{fig:S0hydro0.76}.
Here $S_{min}$ denotes the lowest value of
structure factor $S$. The full, dashed and long-dashed curves refer to
$S_{nn}$, $S_{ll}$ and $S_{\perp}$ respectively.

\end{enumerate}

\begin{center}
\begin{figure}[h]
\vspace{-0.5cm}
\epsfig{figure=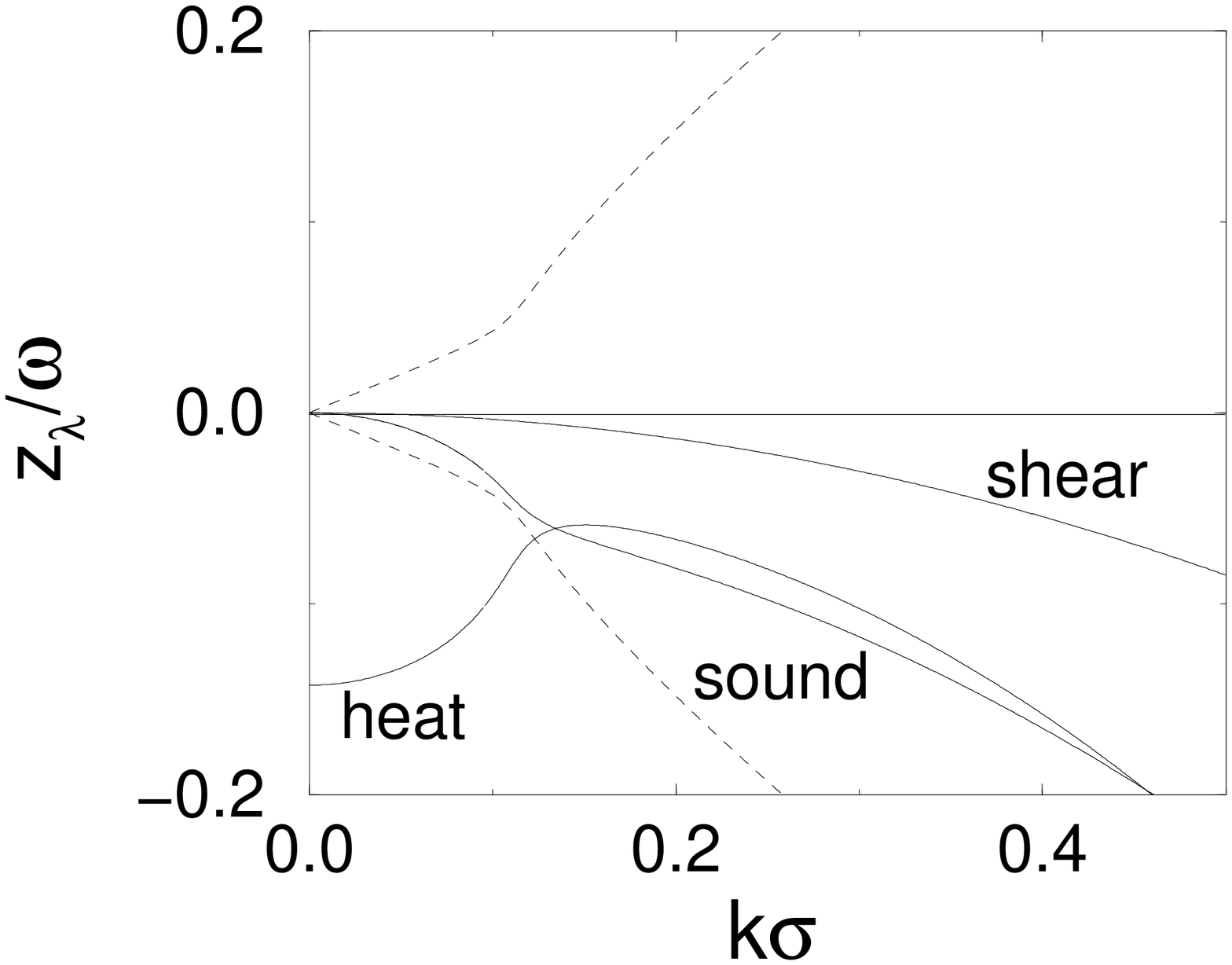,width=13cm,angle=0}
\caption{\label{fig:Stheo}}
\end{figure}
\end{center}

\begin{center}
\begin{figure}[h]
\vspace{-1cm}
\epsfig{figure=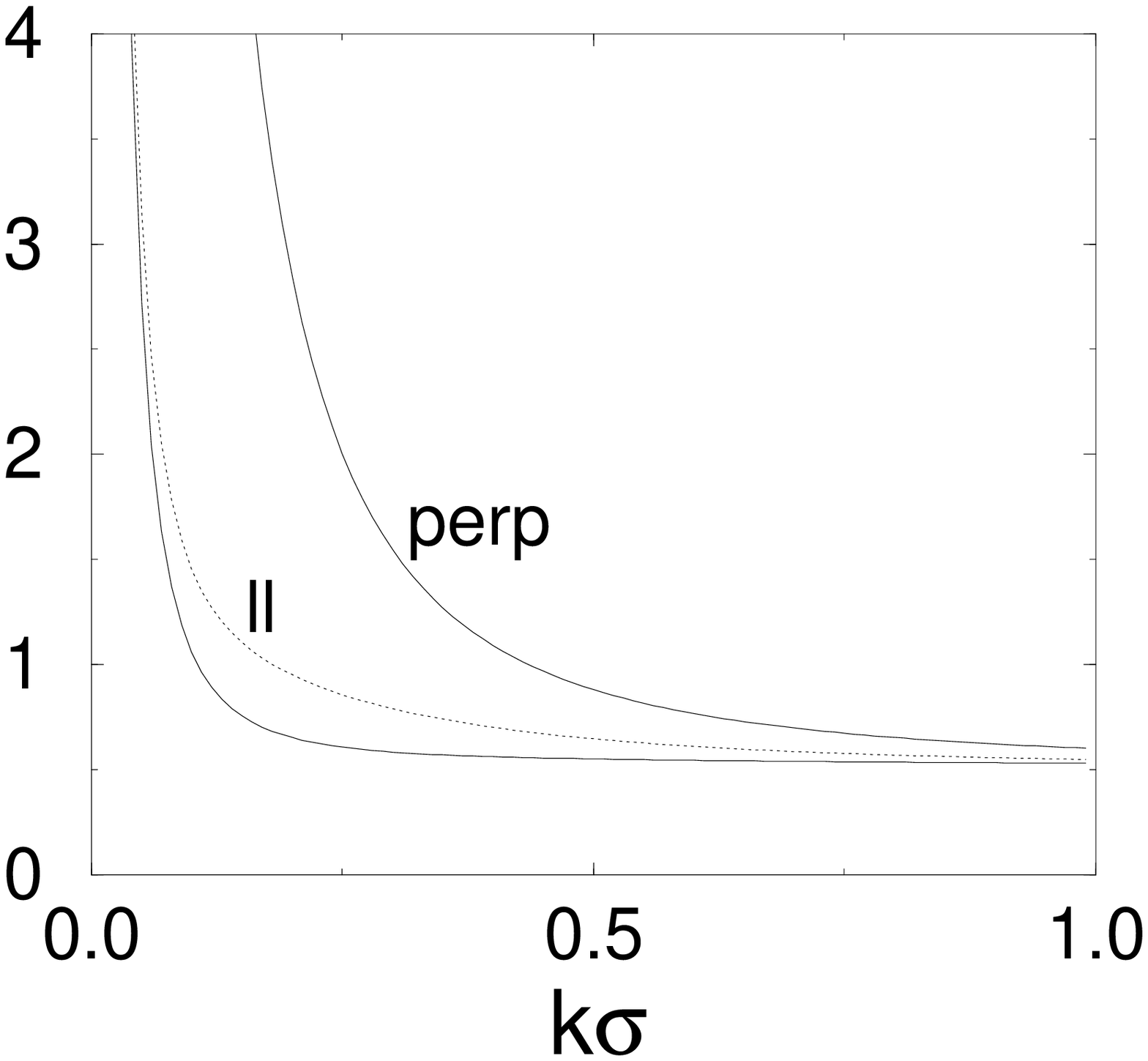,width=13cm,angle=0}
\caption{\label{fig:int-ext}}
\end{figure}
\end{center}

\begin{center}
\begin{figure}[h]
\vspace{-1cm}
\epsfig{figure=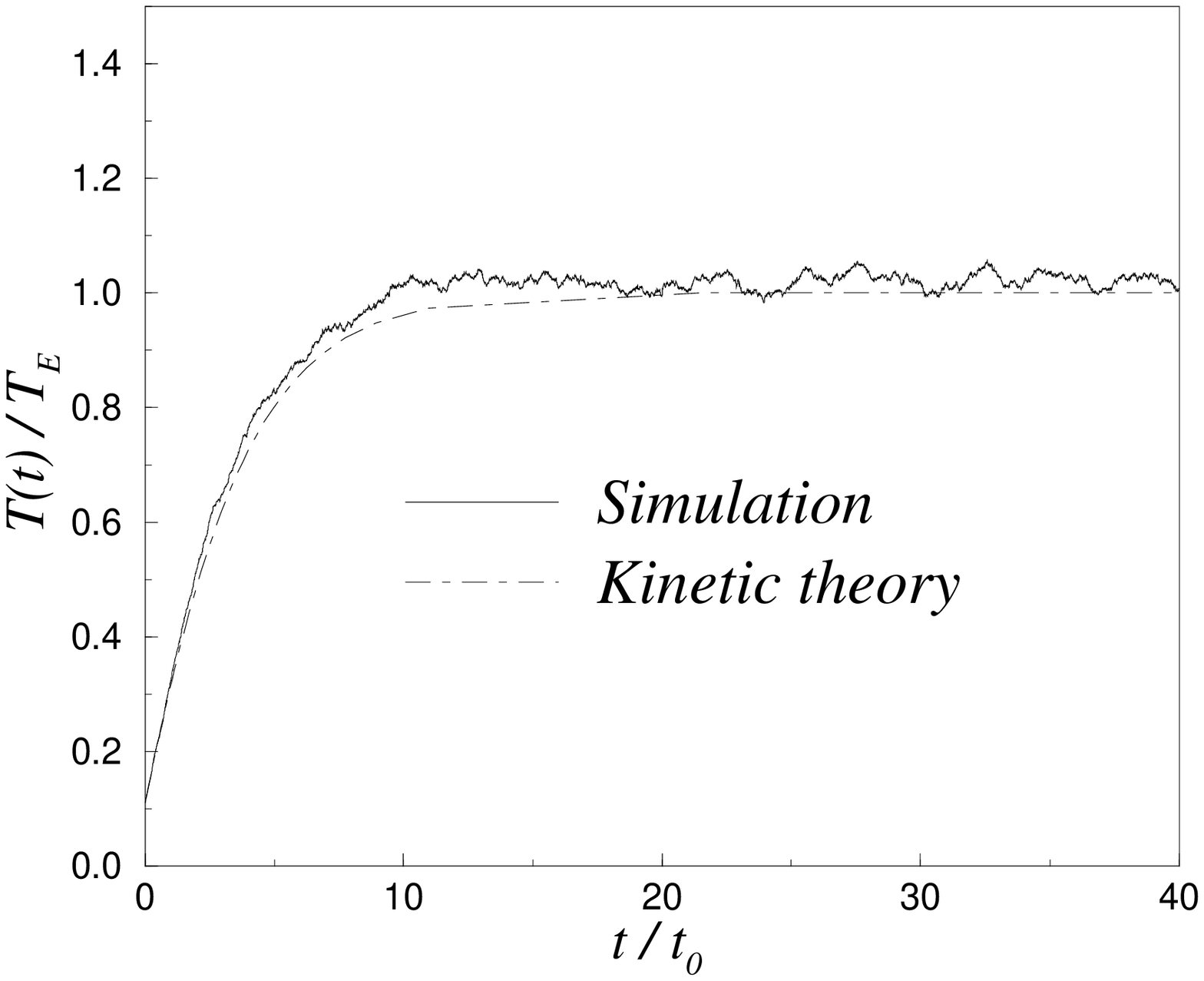,width=13cm,angle=0}
\caption{\label{fig:Tdet}}
\end{figure}
\end{center}

\begin{center}
\begin{figure}[h]
\vspace{-1cm}
\epsfig{figure=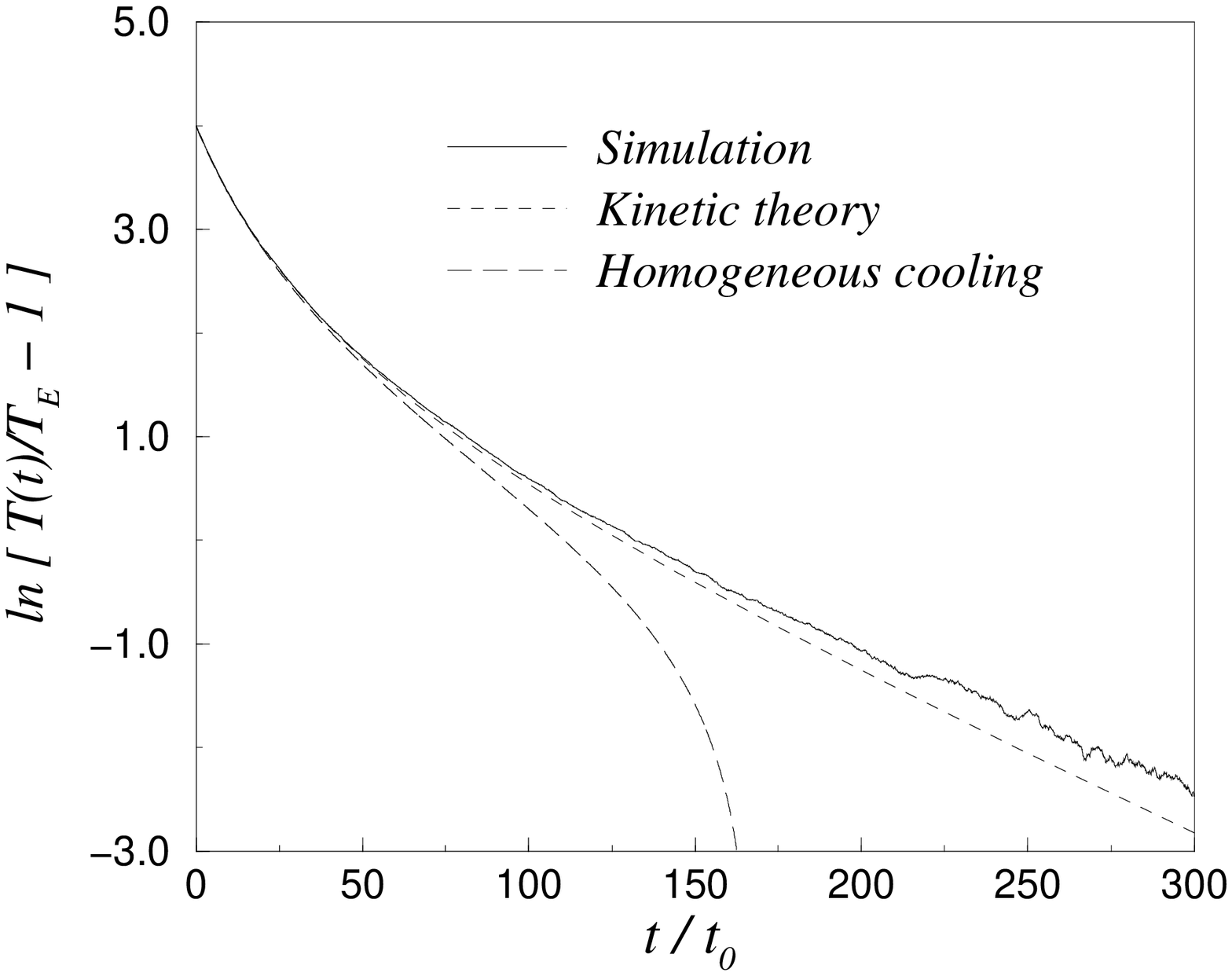,width=13cm,angle=0}
\caption{\label{fig:Tdetlog}}
\end{figure}
\end{center}

\begin{center}
\begin{figure}[h]
\vspace{-1cm}
\epsfig{figure=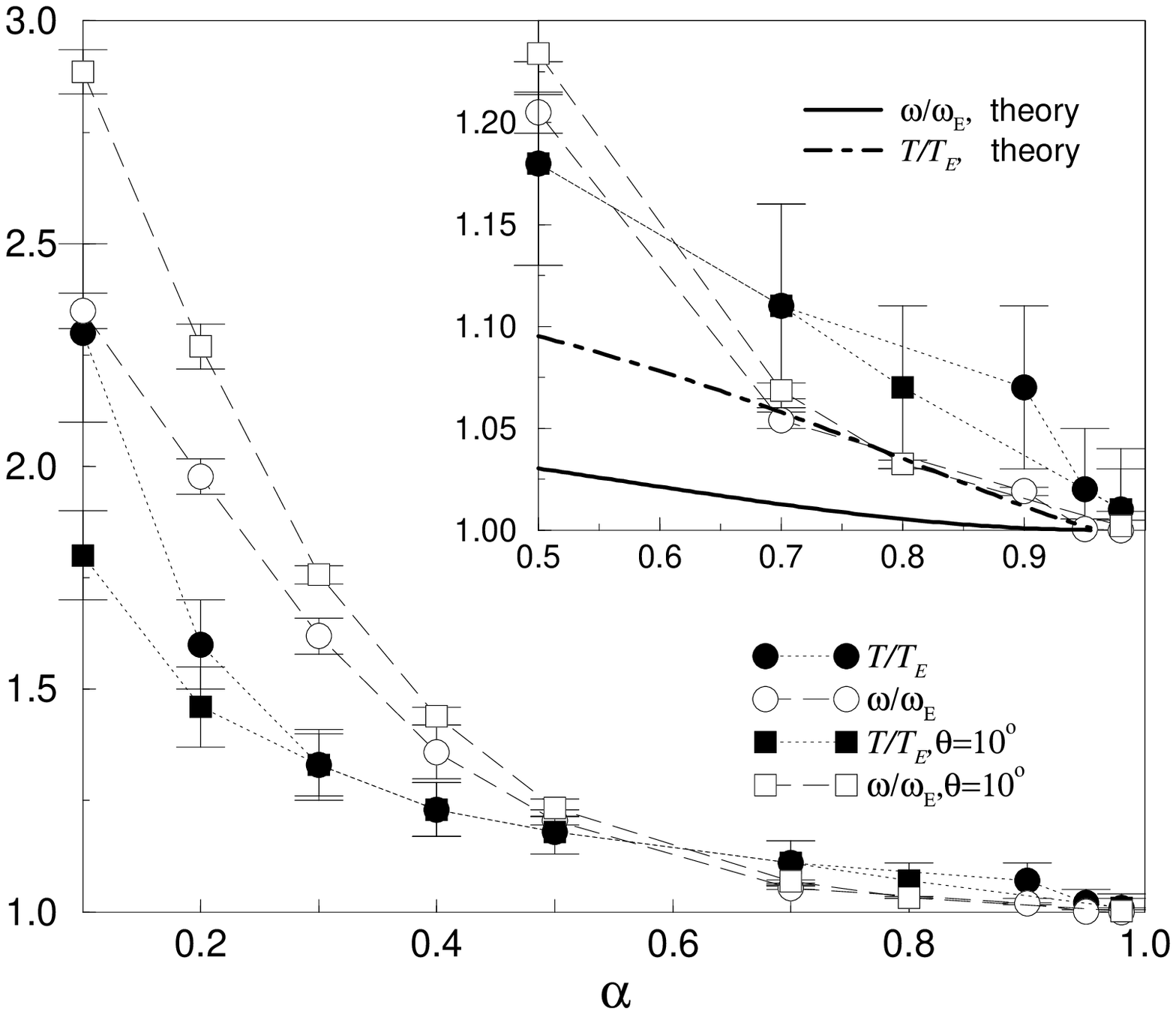,width=13cm,angle=0}
\caption{\label{fig:scale}}
\end{figure}
\end{center}

\begin{center}
\begin{figure}[h]
\vspace{-1cm}
\epsfig{figure=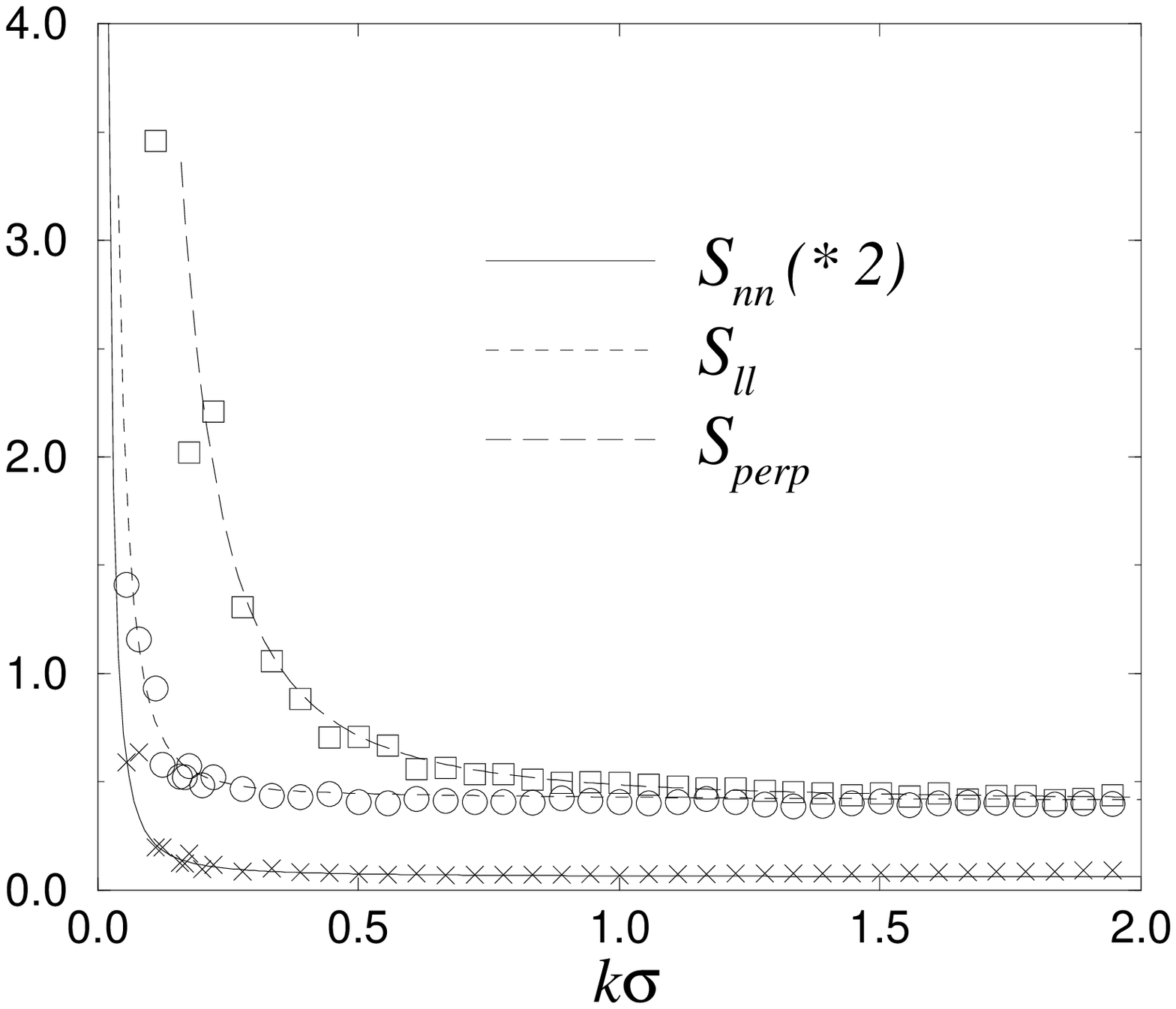,width=13cm,angle=0}
\caption{\label{fig:S0.92}}
\end{figure}
\end{center}

\begin{center}
\begin{figure}[h]
\vspace{-1cm}
\epsfig{figure=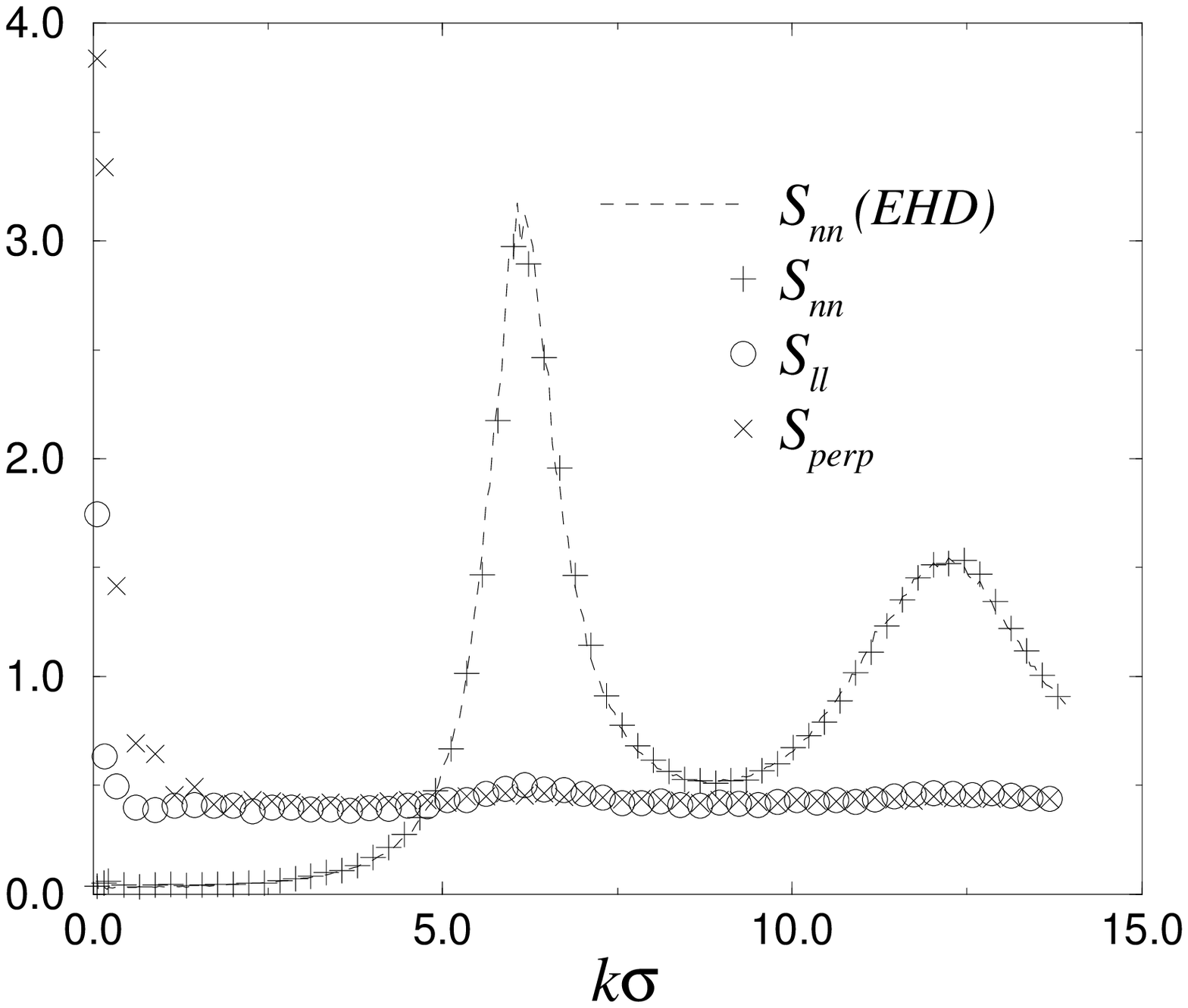,width=13cm,angle=0}
\caption{\label{fig:S0.92_ehd}}
\end{figure}
\end{center}

\begin{center}
\begin{figure}[h]
\vspace{-1cm}
\epsfig{figure=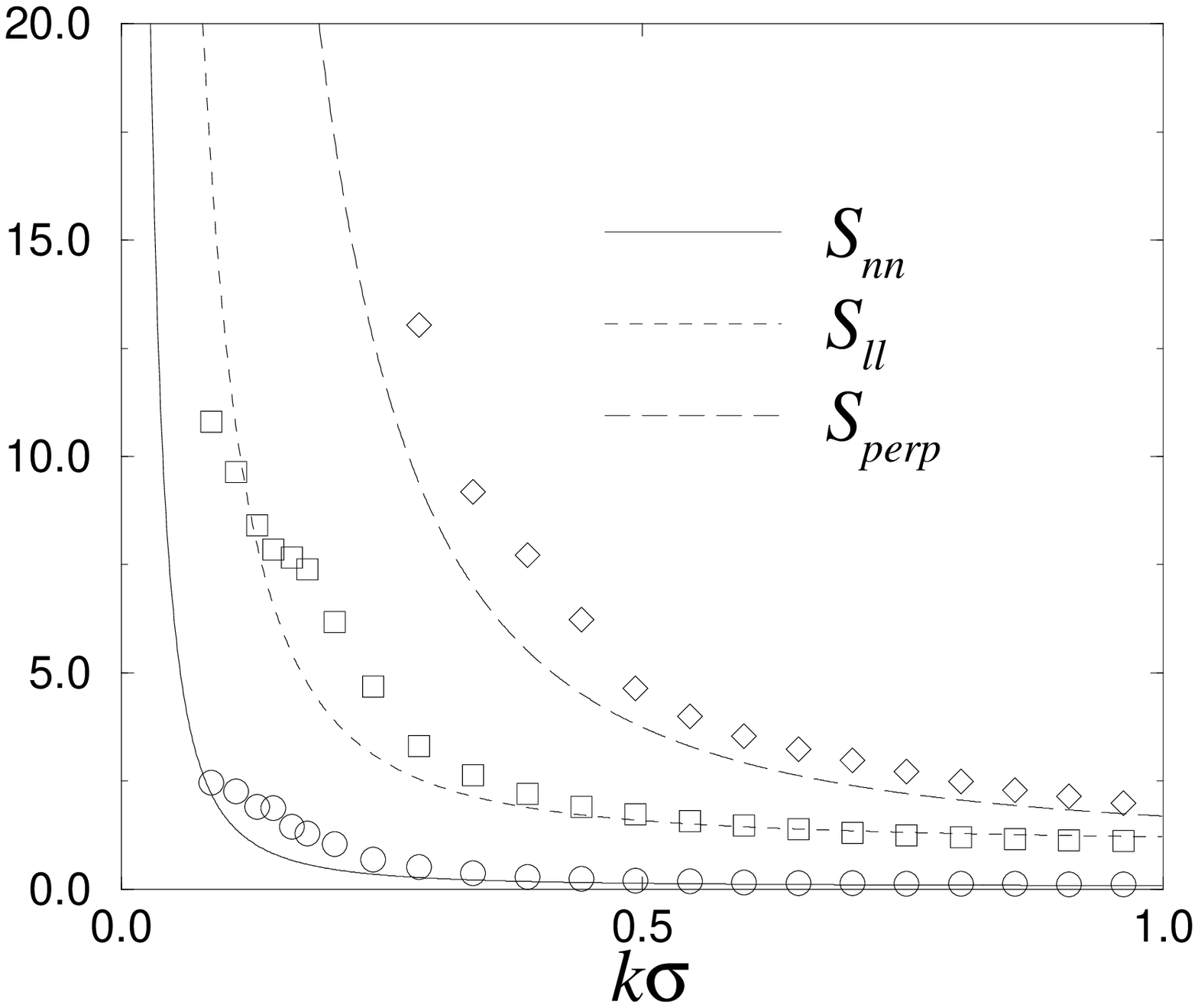,width=13cm,angle=0}
\caption{\label{fig:S0.6hydro}}
\end{figure}
\end{center}

\begin{center}
\begin{figure}[h]
\vspace{-1cm}
\epsfig{figure=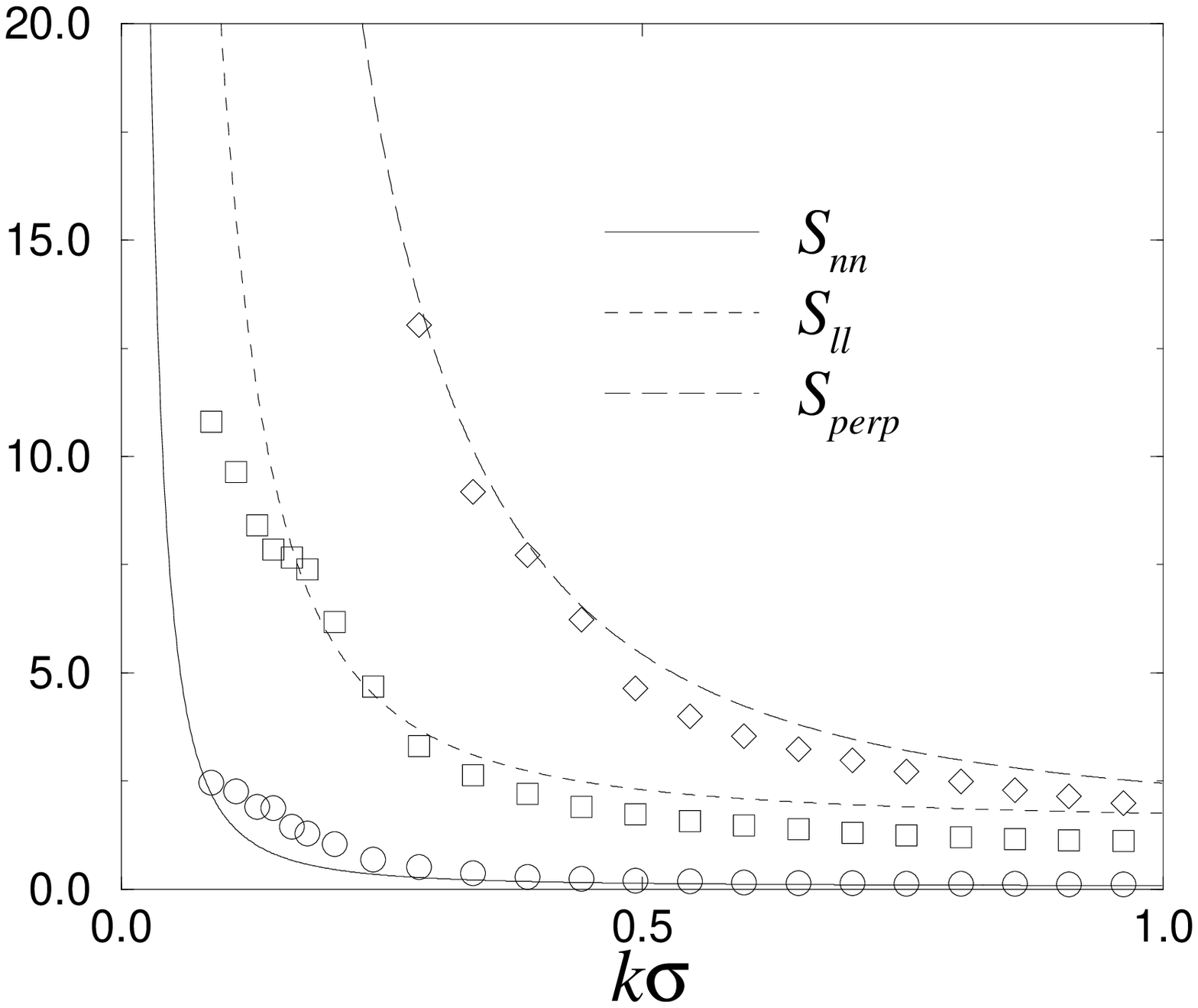,width=13cm,angle=0}
\caption{\label{fig:S0.6hydro1.45}}
\end{figure}
\end{center}

\begin{center}
\begin{figure}[h]
\vspace{-1cm}
\epsfig{figure=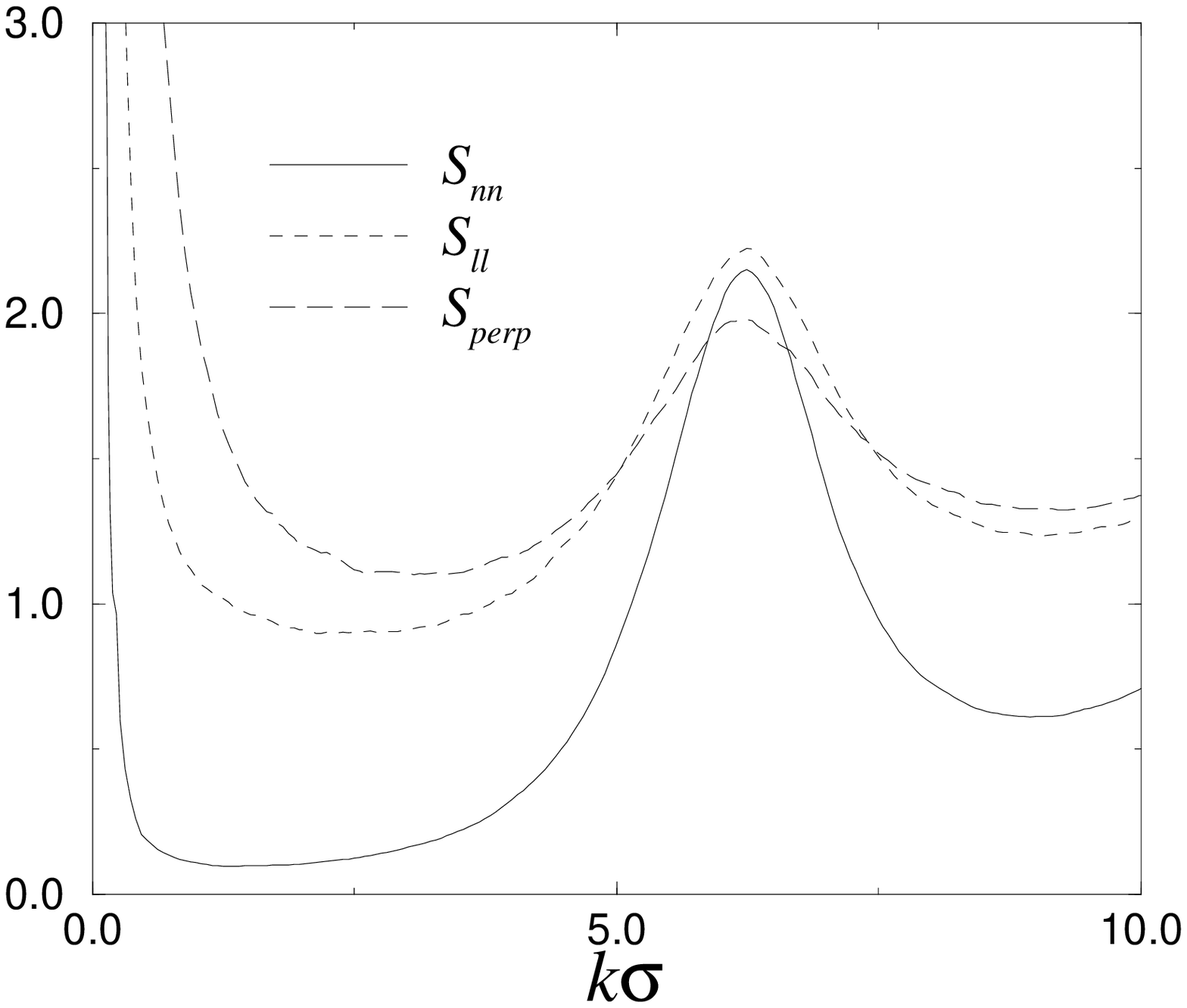,width=13cm,angle=0}
\caption{\label{fig:S0.6tout}}
\end{figure}
\end{center}

\begin{center}
\begin{figure}[h]
\vspace{-1cm}
\epsfig{figure=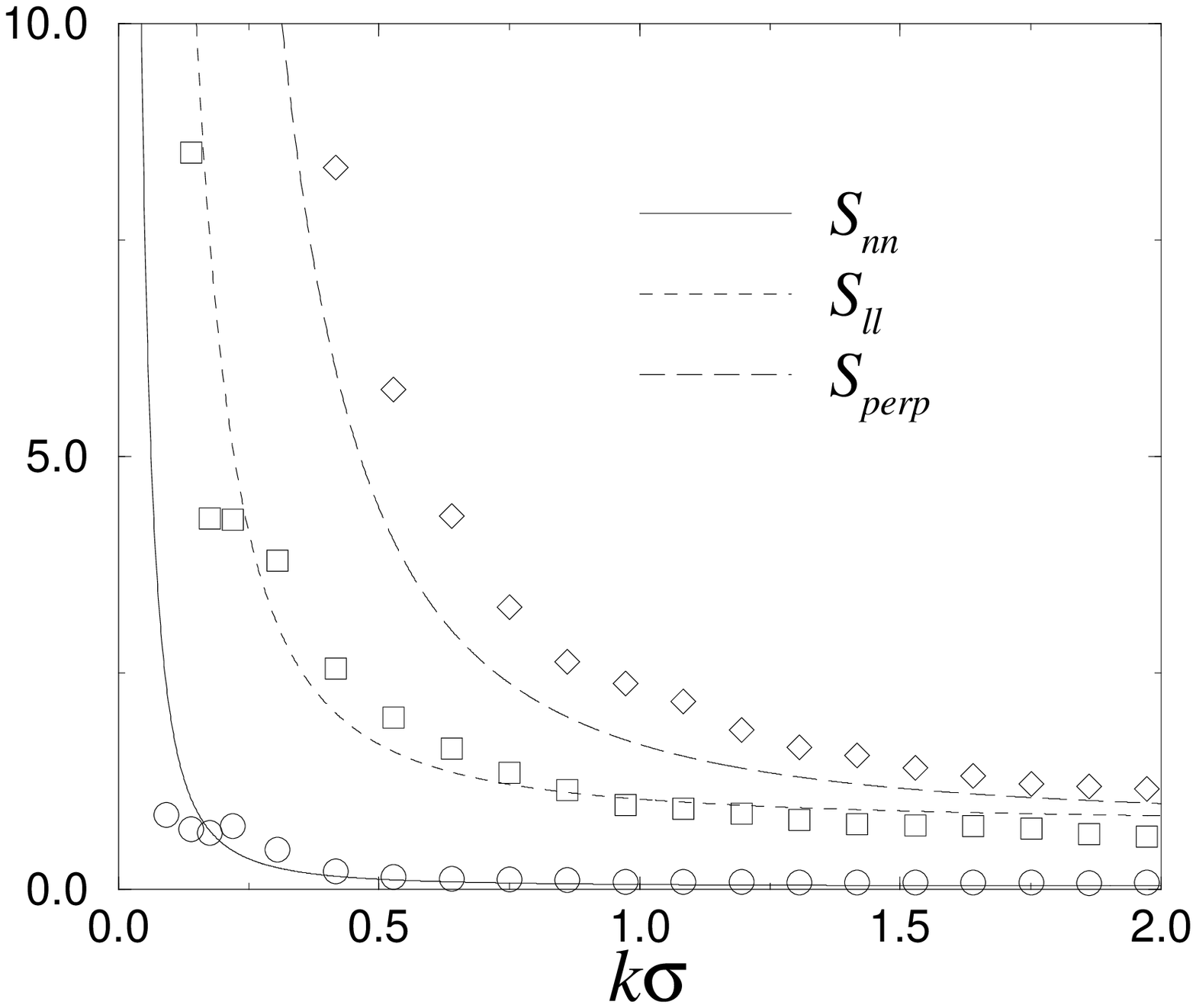,width=13cm,angle=0}
\caption{\label{fig:S0hydro0.76}}
\end{figure}
\end{center}

\begin{center}
\begin{figure}[h]
\vspace{-1cm}
\epsfig{figure=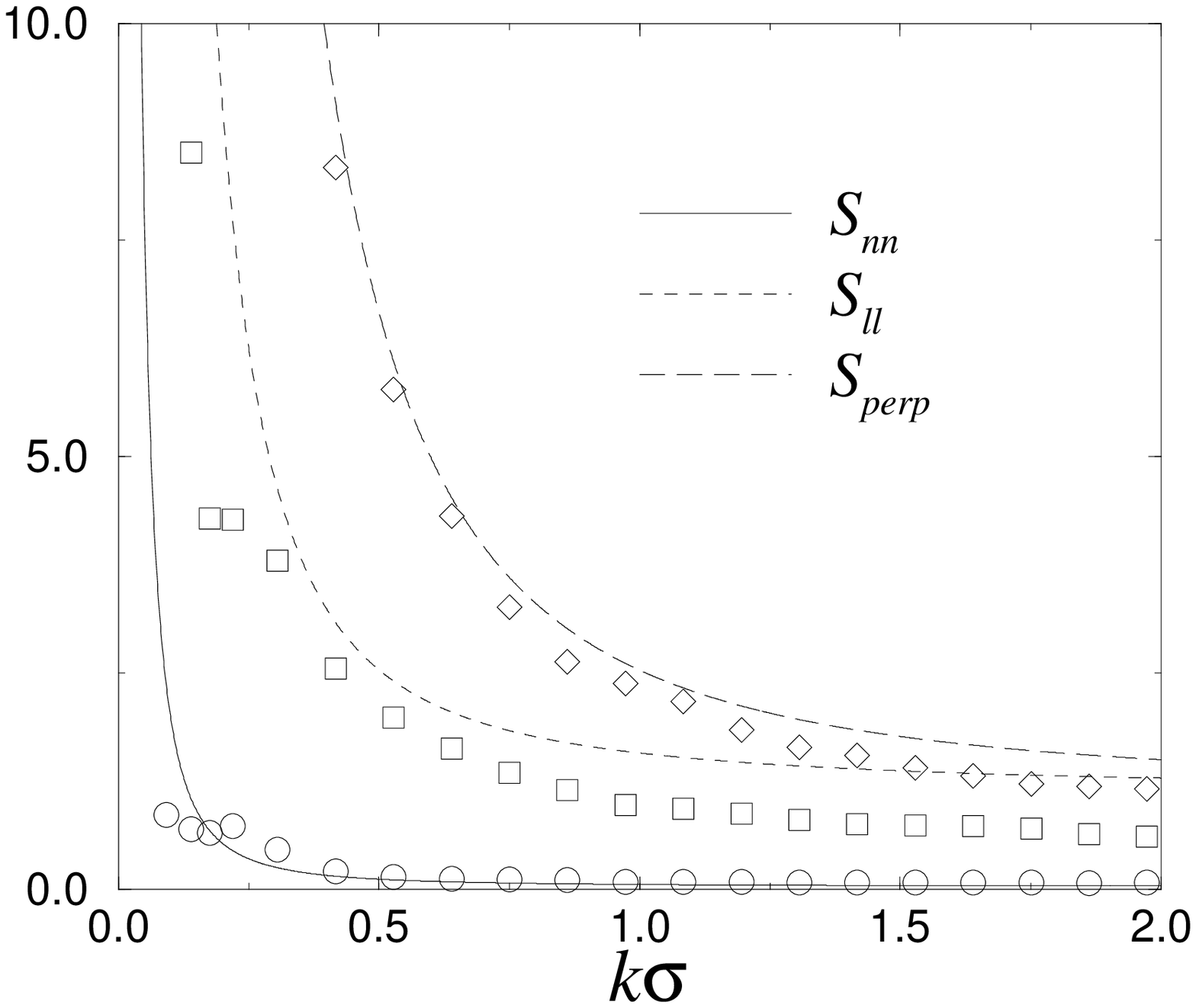,width=13cm,angle=0}
\caption{\label{fig:S0hydro}}
\end{figure}
\end{center}

\begin{center}
\begin{figure}[h]
\vspace{-1cm}
\epsfig{figure=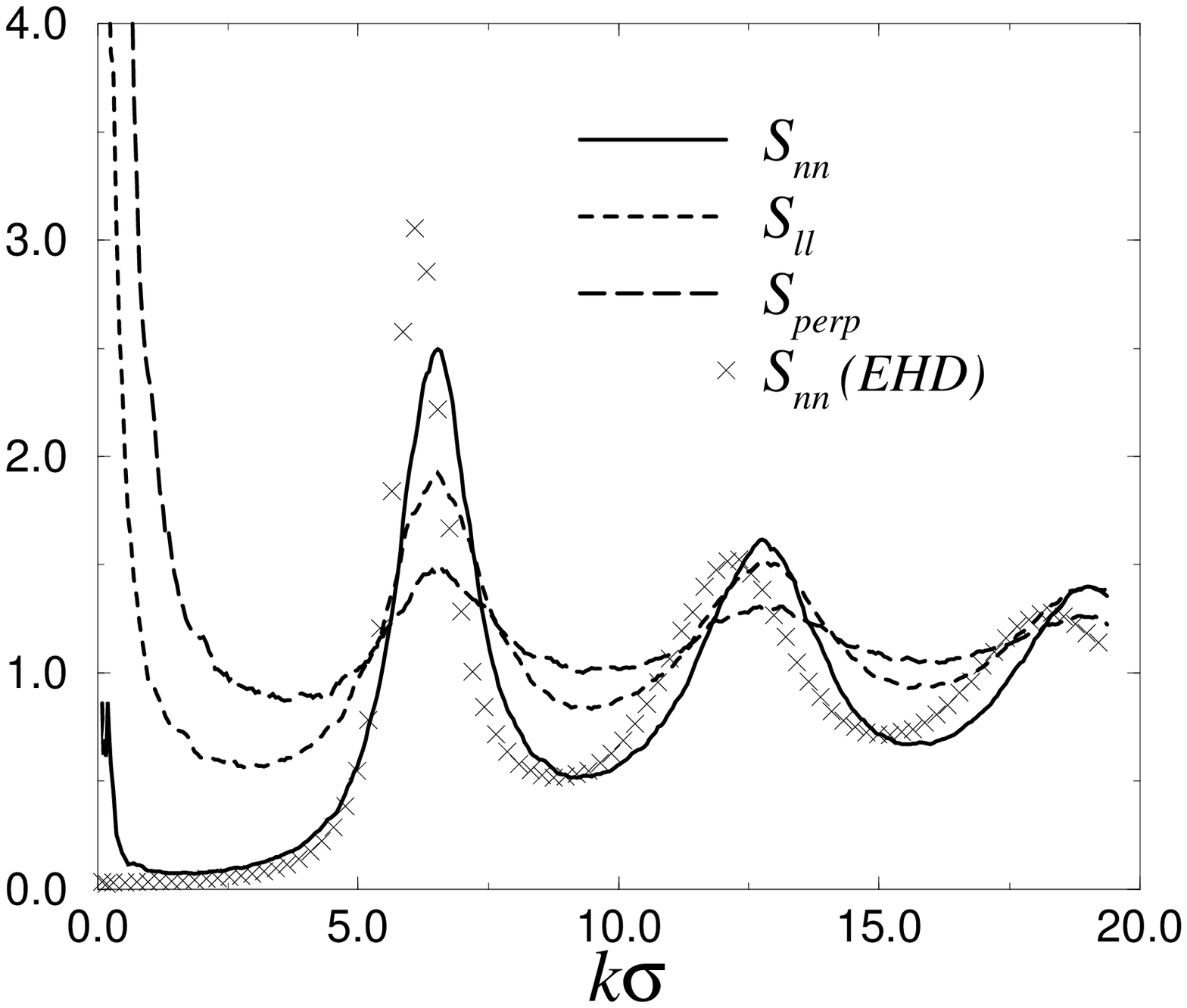,width=13cm,angle=0}
\caption{\label{fig:S0tout}}
\end{figure}
\end{center}

\begin{center}
\begin{figure}[h]
\vspace{-1cm}
\epsfig{figure=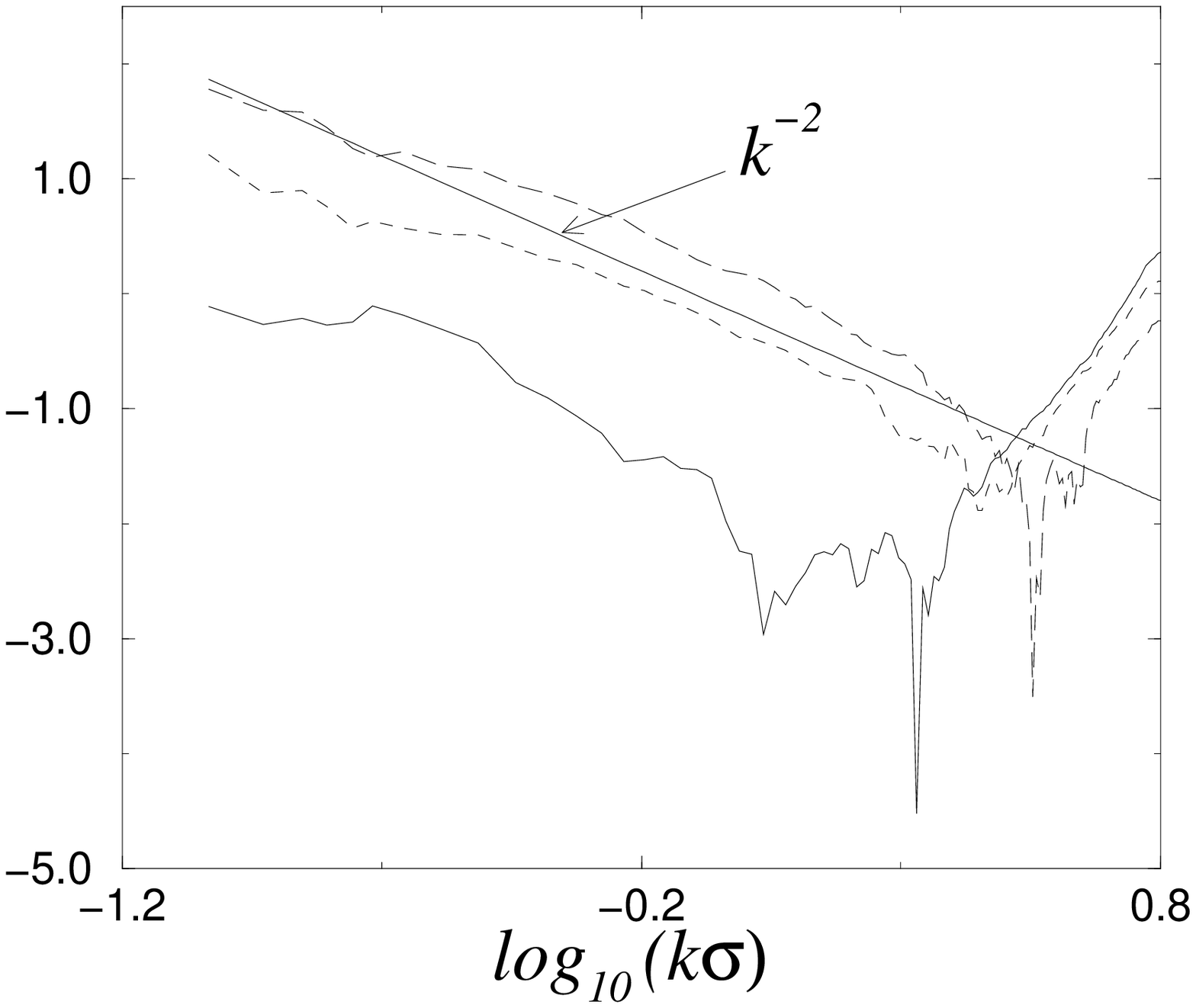,width=13cm,angle=0}
\caption{\label{fig:S0log}}
\end{figure}
\end{center}

\end{document}